# HpC: A Calculus for Hybrid and Mobile Systems – Full Version

XIONG XU*, National Key Laboratory of Space Integrated Information System, Institute of Software, Chinese Academy of Sciences, China
JEAN-PIERRE TALPIN*, INRIA, France
SHULING WANG, National Key Laboratory of Space Integrated Information System, Institute of Software, Chinese Academy of Sciences, China
HAO WU, KLSS (CAS) and SKLCS, Institute of Software, Chinese Academy of Sciences, UCAS, China
BOHUA ZHAN, Huawei Technologies Co., Ltd., China
XINXIN LIU, KLSS (CAS) and SKLCS, Institute of Software, Chinese Academy of Sciences, China
NAIJUN ZHAN**, School of Computer Science, Peking University, China

Networked cybernetic and physical systems of the Internet of Things (IoT) immerse civilian and industrial infrastructures into an interconnected and dynamic web of hybrid and mobile devices. The key feature of such systems is the hybrid and tight coupling of mobile and pervasive discrete communications in a continuously evolving environment (discrete computations with predominant continuous dynamics). In the aim of ensuring the correctness and reliability of such heterogeneous infrastructures, we introduce the hybrid $\pi$-calculus (**HpC**), to formally capture both mobility, pervasiveness and hybridisation in infrastructures where the network topology and its communicating entities evolve continuously in the physical world. The $\pi$-calculus proposed by Robin Milner et al. is a process calculus that can model mobile communications and computations in a very elegant manner. The **HpC** we propose is a conservative extension of the classical $\pi$-calculus, i.e., the extension is "minimal", and yet describes mobility, time and physics of systems, while allowing to lift all theoretical results (e.g. bisimulation) to the context of that extension. We showcase the **HpC** by considering a realistic handover protocol among mobile devices.

CCS Concepts: • **Theory of computation** → **Process calculi**; **Timed and hybrid models**; *Algebraic language theory*; *Operational semantics*.

Additional Key Words and Phrases: $\pi$-calculus, Hybrid Systems, Internet of Things, Bisimulation

## 1 INTRODUCTION

Today's Internet of Things (IoT) consists of networked cybernetic and physical systems (CPSs) that improve coordination and observation of large infrastructures by sensing and actuating the environment using increasingly complex distributed protocols. Most of such systems consist of heterogeneous endpoints, physical or cybernetic, of highly variable capacities, resources, connectivity, mobility, and security. These endpoints combine hardware-embedded functionalities in complex, distributed, discrete/continuous control loops between small, autonomous and mobile endpoints and large, secure, high-performance servers, often so from individual, sporadic or intermittent interactions between "end-user apps", e.g., smartphones, vehicles, and large infrastructures, e.g., clouds, traffic or satellite networks. When designing networked and mobile CPSs of the IoT, we need to take both the effect of the dynamic change of the network topology (mobility) and the correctness of the hybrid behaviours (continuous evolution, discrete computation and control, and their fusion) of the communicating entities within the network into account. We call such systems *Hybrid and Mobile Systems*, of which the complexity reflects in the following four aspects.

**Mobility** The physical location as well as the communication link of an entity within the network keeps changing, i.e., the communication topology is dynamically changed as private client-server channels are transferred as messages using servicing channels.

**Privacy** The communication within the network usually has privacy requirements. Some entities may form a local group to communicate with each other, but other joining the network may not share with this group.

**Openness** In the runtime of the network, new entities and communication links may be created and, meanwhile, old or obsolete ones may be deleted at any time, i.e., the network is an open and dynamic system.

**Hybridisation** Physical and cybernetic entities in the network perform hybrid behaviour, coupling discrete sensing and actuation protocols and topologically pervasive continuous physical behaviours.

Hybrid and mobile systems are usually safety-critical, for instance, internet-connected vehicles, next generation train networks, air-traffic control, unmanned autonomous systems. Any defect in the design of protocols controlling such distributed systems may lead to catastrophic consequences. Assuring the safety of heterogeneity systems with mobility and dynamicity forms a challenge which must be addressed with formal guarantees as mandated by many certification standards for safety-critical system design. Dealing with the tightly-coupled combination of mobility, privacy, openness, and hybridisation is the most difficult aspect of this challenge. Unfortunately, and to the best of our knowledge, no existing calculus covers the full spectrum and complexity of such infrastructures, despite there commonality in civil and industrial networks.

To address this issue, this article presents a conservative extension of the classical $\pi$-calculus, the hybrid $\pi$-calculus (**HpC**). In the $\pi$-calculus, proposed by Robin Milner et al., channels, variables, and data are all first-class citizens: they are called names. This makes a fundamental difference between the $\pi$-calculus and other process calculi such as CSP (Communicating Sequential Processes) or CCS (Calculus of Communicating Systems). A name can be used as a channel to transfer other names and meanwhile it can be treated as a message transferred by names: a first-class citizen. More details about $\pi$-calculus can be found in [35–38]. The convention that "names are first-class citizens" makes the $\pi$-calculus powerful in characterising mobility. Therefore, we adopt this convention in the **HpC**, where a name can uniformly be used as a channel, a parameter, an ordinary, continuous or physical variable. Thanks to this convention, the **HpC** can model complex networked and mobile CPSs in a very natural and elegant manner, just like the classic $\pi$-calculus for discrete mobile communication systems alone.

*Contributions.* This paper presents the first characterisation of a $\pi$-calculus that has the capability to compositionally orchestrate the interaction of discretely timed processes and continuously timed ordinary differential equations (ODEs). The **HpC** provides a hybrid calculus in which all objects are first-class citizens:

- Discrete channels and continuous variables are treated uniformly;
- Discrete and continuous processes are equal hybrid processes;
- All channels and processes, discrete or continuous, have both meanings in the discrete and continuous operational semantics; and
- Last but not least, the **HpC** is the first calculus to support control-theoretic approximate bisimulation in the presence of dynamic and mobile channels and processes.

This combination forms a process algebra, the **HpC**, to describe protocols amongst hybrid and mobile systems by allowing for the specification of mobility, real-timed continuous evolution and discretely-timed concurrent computations and communications. The lean and simple embedding of discrete and continuous functions in the $\pi$-calculus (there is one proof system for discrete interaction and one for continuous evolution) allows us to define a theory of bisimulation applying to discrete and continuous processes alike, providing the most expressive model for the verification of hybrid and mobile systems. While perhaps surprising, this result is in turn similar to, and an extension of, earlier results in [41], which proved the discrete relative completeness of differential dynamic logic, and in [59], which proved the discrete relative completeness of hybrid Hoare logic. We introduce the real-world case study of a handover protocol for a train control system to demonstrate the benefits of the **HpC** to provide an expressive, sound and precise modelling and verification workflow to model protocols dominated by physics, distribution, real time, dynamicity and mobility, along with a theory of bisimulation to verify them.

*Paper organisation.* After reviewing related works in Section 2, some motivating examples in Section 3 give an overview of the **HpC**, whose formal syntax and operational semantics are specified in Section 4 and Section 5. Then, a theory of strong, weak, and approximate bisimulation for the **HpC** are introduced in Section 6. The case study of a handover protocol of a real-world train control system is demonstrated in Section 7. Section 8 concludes.

## 2 RELATED WORKS

An important research field and corpus of related works addresses the modelling and verification of combined continuous and discrete dynamics, also called hybrid systems. It can be roughly classified into the following categories:

- Graphical modelling tools with simulation engines (Simulink [33, 34], Ptolemy, Modelica), allow to model closed-loop systems using diagrams, verify and validate them by simulation.
- Automata-based modelling and model-checking based on reachable sets computation abstractly model systems using hybrid automata [1, 23]. Hybrid automata extend finite-state automata by attaching ordinary differential equations (ODEs) to model continuous evolution in each state of the system. Verification is conducted by computing reachable sets possibly with abstraction and approximation [13, 17, 18, 27];
- Process algebra-like modelling together with theorem proving, by which a system is modelled as a process-algebraic term, and verification is done by theorem proving, e.g., differential dynamical logic [40] and hybrid Hoare logic [31, 59].
- Extended Event-B framework, based on refinement and automated proof to model and verify hybrid systems, for instance, core hybrid Event-B [6–8] and Event-B hybridisation [16].

Unlike the aforementioned approaches, advanced concurrency models such as the $\pi$-calculus [35] and its variants do support the expression of mobility and dynamic communication topology. With these process-algebraic methods, verification is conducted using several forms of bisimulation such as strong, weak, and branching bisimulation [38, 50]. None of the aforementioned approaches to the modelling and verification of hybrid systems address mobility and dynamicity.

### 2.1 Process Algebras for Timed and Hybrid Systems

The notion of Hybrid Process Algebra (HyPA) was first proposed in [15] and it is an extension of the Algebra of Communicating Processes (ACP) to model hybrid systems. HyPA processes were then linearised in [49], where a corresponding simulation tool was provided. Hybrid $\chi$ [47] is an extension of the simulation language $\chi$ to model hybrid systems. Hybrid $\chi$ combines the theories of process algebra and hybrid automata and can model the dynamic behaviour of systems. Reference [48] defined the methods that transform hybrid $\chi$ to other formal models that describe hybrid systems. The work of [10] combined the process algebra with timing [5], propositional signals [4], and a process algebra that can describe the interleaving of discrete actions and continuous evolution.

A process algebra combining time and mobility, named TiMo (Timed Mobility), was proposed in [14] and it was then extended to real-time systems as rTiMo [2]. However, rTiMo only supports local communication, while in the real-world application, entities can communicate remotely. Thus, a new model BigrTiMo [56], the combination of Bigraph model and rTiMo, was present. Compared to the current calculus for mobile distributed systems, BigrTiMo can not only capture mobility, but also describe space and time constraints. Recently, a timed calculus with mobility, called MTCWN [55], was proposed to capture mobility in wireless networks.

Of inspiration to the definition of the **HpC** is He's works on Hybrid CSP (HCSP) [22, 60] which provide a uniform characterisation of discrete processes and continuous behaviours in Hoare's CSP [24]. However, it does not support mobility, i.e., the dynamic introduction and transmission of channel names as first-class objects. The classical $\pi$-calculus supports mobility that way, but does depict the continuous evolution of CPSs. The $\Phi$-calculus, the first extension of $\pi$-calculus to model hybrid systems, was proposed in [45], and a corresponding verification framework based on spatial logic was given in [44]. However, the issue of mobility is not the concern of the calculus, and the tight coupling of hybrid behaviour and mobile communication is also not addressed.

The session $\pi$-calculus [46] is an extension of the $\pi$-calculus that can model communication protocols between processes in a more structured manner. The calculus enjoys a powerful multiparty session type system by which properties like deadlock-freedom and liveness can be proved. However, it cannot model continuous behaviour. Another calculus of sessions is the multiparty motion session calculus [32]. It is equipped with continuous-time motion primitives to describe affine approximations of continuous agents behaviours. It serves as a modelling language for programming and schedule the coordination of multiple robots. However, it does not support general forms of ODEs or mobility, i.e., the topology of communication never changes.

### 2.2  Process Algebras for the Internet of Things

There are some works based on process calculi for the modelling and verification of IoT. For example, the IoT-calculus [29] is the first process calculus for IoT. It can capture some basic features of IoT networks, such as the partial topology of communications and the interaction between sensors, actuators and computing processes to provide useful services to the end user. IoT-LySa [11] is a process calculus endowed with a static analysis that tracks the provenance and the route of IoT data, and detects how they affect the behaviour of smart objects. However, neither IoT-calculus nor IoT-LySa can describe timed behaviour. A process calculus of IoT systems, called CaIT, is proposed in [30], where timed behaviours, with desirable time, consistency and fairness properties, are supported. Recently, a secure mobile real-time process calculus for specifying and reasoning about IoT systems, called SMrCaIT [12], was proposed. It supports not only value-passing communication but also name-passing communication, and additionally, it can strictly separate process actions and mobility modelling by providing parametric mobility models. However, SMrCaIT cannot describe the continuous evolution of entities in physical world.

### 2.3  Verification of Hybrid Systems

A lot of existing works on verification of hybrid systems use reachability-analysis and theorem proving methods. Most of the works on reachability analysis model hybrid systems use hybrid automata, which are verified by computing reachable sets [1]. On the theorem proving side, differential dynamic logic [40, 42] reasons about hybrid systems using an extension of dynamic logic to include continuous evolution. HCSP [22] is an extension of CSP that includes continuous evolution and interruption by communication. Proof systems based on Hoare logic have been proposed for HCSP [21, 31, 59]. In this paper, we conduct the verification of **HpC** processes based on the notion of bisimulation, which covers the behaviour of discrete communication, mobility of names, dynamicity of processes, and of continuously timed functions.

There have been some work on simulation and approximate simulation relations between hybrid systems. Due to the presence of ODEs, it is not possible to maintain an exact simulation relation between hybrid systems or between a hybrid system and a discrete program executable on a computer. Hence, the concept of approximate bisimulation is proposed [20] to allow errors up to a bounded distance between states of the two systems. The distance between two states is usually defined as the maximum of Euclidean distances of values of variables under states, which is actually one special case of the quantitative bisimulation proposed in [19, 28]. The work [25] proposes the use of two parameters on both observable states and transition labels as precisions. Using a similar notion of approximate bisimulation with two parameters, the work [58] proved correctness of code generation from HCSP models into SystemC [58] and then into C [53], in the sense that the original model and the discrete code are approximately bisimilar.

## 3 OVERVIEW: MOTIVATING EXAMPLES

We introduce some preliminary examples to illustrate how continuous behaviours are represented in the **HpC**, starting with continuous time, and then highlight more of its features using more elaborate ones.

$$\texttt{BigBen} \triangleq \{0 \mid \dot{c} = 1, \{\overline{c}\}\}$$

The process named BigBen has a free name $c$ denoting the clock (highlighted in blue for clarity). Intuitively, the implementation of BigBen only interacts with the outside world using the (discrete) communication channel $c$. The evolution of BigBen is continuously defined by an ordinary differential equation (ODE) $\dot{c} = 1$ with the initial condition 0, where $\dot{c}$ is the abbreviation of the derivative $\frac{dc}{dt}$ and $\dot{c} = 1$ means $c$ increases linearly (with respect to time). The set $\{\overline{c}\}$ defines the *interface* of the ODE. The notation $\overline{c}$ is an output action, indicating that the real-time value of the clock $c$ can be sensed (read) at any time during its evolution. However, the clock of BigBen cannot be actuated (written). This would have to be specified by an additional input action $c$ in the interface, i.e., $\{\overline{c}, c\}$, allowing one to reset BigBen with a new value of $c$. BigBen can serve as a global clock that can be sensed everywhere in the scope of $c$. We can, on the contrary, use $c$ as a private name to model the side effect of a timing constraint, as follows:

$$\texttt{wait}(d) \triangleq (\nu c)\{0 \mid \dot{c} = 1 \ \& \ c < d\}$$

Process wait is parameterised by $d$, to denote the duration and the scope of $c$ (private and localised by $\nu$). No one in the environment can read or write its value (the interface of the ODE is empty). The constraint $c < d$ means that the evolution of the ODE terminates once the value of $c$ reaches $d$. It is called the *boundary condition* of the ODE. Concretely, one may depict the evolution of process wait(d) by the following transition:

$$\texttt{wait}(\texttt{d}).\texttt{P} \xrightarrow{d} \texttt{P}$$

In the reminder, we will reuse process wait to express time delays. Now, let us consider a process observing time from BigBen repeatedly. The process named Observer recursively (modality $\mu$) reads time $t$ from channel $c$, performs some time-sensitive task Task($t$) and then waits for two time units before scheduling the next run. The Observer can interact with BigBen periodically by sharing channel $c$. This interaction can be modelled by the parallel composition BigBen ∥ Observer.

$$\texttt{Observer} \triangleq \mu x.c(t).\texttt{Task}(t).\texttt{wait}(2).\overline{x}$$

REMARK 1. *Notice that name $c$ in* BigBen *is a physical variable of the ODE $\dot{c} = 1$. Meanwhile, it is used as an output channel (a sensor) in its analog-digital interface $\{\overline{c}\}$ with process* Observer. *The clock $c$ evolves continuously according to $\dot{c} = 1$. During its evolution,* BigBen *can provide a real-time value to the observer using the discrete output channel $\overline{c}$, whenever it is requested to. In this sense, $c$ could be called a "continuous channel". In the* **HpC**, *however, we do not distinguish between physical variables and channels. They are all "first-class citizens" following the convention of the $\pi$-calculus.*

### 3.1 Bouncing Ball

Now, let us consider a classic: a ball falling from an initial height of 5m with initial velocity 0m/s. The ball falls with a dynamics given by two ODEs $\dot{h} = v$, its height, and $\dot{v} = -9.8$ m/s$^2$, its velocity. The behaviour of the ball is modelled by the following process:

$$\texttt{Ball} \triangleq \left\{ \begin{array}{c|ccc} 5 & \dot{h} & = & v \\ 0 & \dot{v} & = & -9.8 \end{array}, \{v, \overline{v}\} \right\}$$

The interface $\{v, \overline{v}\}$ indicates that the environment can sense ($\overline{v}$) and actuate ($v$) the velocity of the ball. Once the ball hits the ground, its velocity becomes discontinuous: it is reversed. We assume that the collision of the ball with the ground is inelastic and that its velocity is reversed by a factor of $-0.8$, i.e., the ball bounces back and looses 0.2 energy. The effect of hitting the ground implies:

$$\texttt{Ground} \triangleq \mu x.(\nu c)\{0 \mid \dot{c} = 1 \ \& \ h > 0 \lor v \geq 0\}.v(v_0).\overline{v}\langle -0.8v_0 \rangle.\overline{x}$$

where $c$ is the local time on the ground. It means that the ground waits until the elevation $h$ of the ball reaches 0 ($h \leq 0$) with negative velocity ($v < 0$), i.e., the boundary condition $h > 0 \lor v \geq 0$ becomes **false**. When that happens, it reads the velocity of the ball by the input action $v(v_0)$. The discrete variable $v_0$ stores the value of the velocity and the ground inverts it to $-0.8 \times v_0$ by performing the output action $\overline{v}\langle -0.8v_0 \rangle$. The bouncing ball system can be defined by the parallel composition Ball∥Ground, which causes a Zeno-behaviour: it generates an infinite execution but the sum of the durations of all the timed actions in the execution is bounded by a constant.

### 3.2 Mobile Vehicle

At last, we introduce an example reflecting the capability of the **HpC** in modelling mobility, i.e., the migration of channel and physical variable names as messages through servicing channels between processes. We consider a vehicle moving back and forth between two base stations, which control the movement of the vehicle alternately. The behaviour of the Vehicle can be modelled as follows:

$$\texttt{Vehicle} \triangleq (\nu p, v, a)\overline{b_1}\langle p, v, a \rangle. \left\{ \begin{array}{c|ccc} 0 & \dot{p} & = & v \\ 0 & \dot{v} & = & a \\ 0 & \dot{a} & = & 0 \end{array}, \{\overline{p}, \overline{v}, a\} \right\}$$

Modality $\nu$ creates names $p$, $v$, and $a$ to denote its position, velocity, and acceleration, respectively. These names are private to the Vehicle, which then sends them using the shared channel $b_1$ to notify base station Base$_1$ that it may sense and actuate the Vehicle using these names. Once synchronised, the vehicle behaves according to the set of ODEs $\{\dot{p} = v, \dot{v} = a, \dot{a} = 0\}$, where $p$, $v$, and $a$ are set to 0 initially. The set $\{\overline{p}, \overline{v}, a\}$ defines the interface of the ODEs: the output actions $\overline{p}$ and $\overline{v}$ denote that the real-time values of position ($p$) and velocity ($v$) of the vehicle can be sensed (read); the input action $a$ indicates that the value of the acceleration ($a$) of the vehicle can be actuated (written) by the environment interacting with the ODEs.

$$\texttt{Base}_1 \triangleq \ !b_1(p, v, a).p(p_0). \begin{pmatrix} [p_0 < 100].v(v_0).\overline{a}\langle f_1(p_0, v_0)\rangle.\texttt{wait}(1).\overline{b_1}\langle p, v, a\rangle \\ + \\ [p_0 \geq 100].\overline{b_2}\langle p, v, a\rangle \end{pmatrix}$$

Notice that input actions $p$ and $v$ are not in the interface, indicating that the position ($p$) and velocity ($v$) of the vehicle cannot be modified by the environment. The behaviour of base station $b_1$ uses the operator + to denote a non-deterministic choice between two conducts and replication (!) to model mutual recursion (between stations 1 and 2). Replication !P is equivalent to P||!P. Intuitively, it is the parallel composition of infinite Ps. When an instance of P is forked, it receives the actual parameters ($p, v, a$) on channel $b_1$ and then forks a process to read the real-time position $p_0$ of the vehicle by its name $p$, i.e., $p(p_0)$. If the vehicle is still within the domain of the base station ($p_0 < 100$), then (1) the base station continues to read the velocity by its name $v$, i.e., $v(v_0)$, (2) adjust the acceleration ($a$) according to the sensed velocity ($v_0$) and position ($p_0$) of the vehicle by an algorithm $f_1$, making the vehicle move forwards, i.e., $\overline{a}\langle f_1(p_0, v_0)\rangle$, (3) wait for 1 second, i.e., wait(1), and (4) repeats the procedure by rebooting itself, i.e., $\overline{b_1}\langle p, v, a\rangle$. Otherwise, if the vehicle has moved out of range of the station ($p_0 \geq 100$), the station contacts its neighbour to take control over the vehicle by sending it the physical variables ($p$, $v$, and $a$) of the vehicle over channel $b_2$. Base$_1$ will then wait until being invoked again by the other station Base$_2$ over channel $b_1$.

Symmetrically, the behaviour of the second base station is modelled by the matching replication. Function $f_2$ orders the vehicle to move back to the first station until the position of the vehicle reduces to 10. When that happens, the first base station is invoked by $\overline{b_1}\langle p, v, a\rangle$ to take control back on the vehicle again. The system can be modelled using parallel composition Vehicle||Base$_1$||Base$_2$, allowing the vehicle to move back and forth between the two base stations.

$$\text{Base}_2 \triangleq \; !b_2(p,v,a).p(p_0).\begin{pmatrix}[p_0 > 10].v(v_0).\overline{a}\langle f_2(p_0,v_0)\rangle.\texttt{wait}(2).\overline{b_2}\langle p,v,a\rangle \\ + \\ [p_0 \leq 10].\overline{b_1}\langle p,v,a\rangle\end{pmatrix}$$

## 4 THE HYBRID $\pi$-CALCULUS

In this section, we introduce the syntax of the **HpC**, along with the notations and the structural congruence laws that will be used throughout the paper.

### 4.1 Names and Expressions

We assume an enumerable set of *names* $\mathcal{N}$, e.g. $x, y, z, \cdots$, possibly with superscript or subscript. Unlike discrete channels, which carry values, continuous physical variables are designated by a subset of names $\mathcal{V} \subseteq \mathcal{N}$ to denote functions from time to values, usually noted $v, u, w$, etc. Expressions $e$ are defined by the grammar $e ::= c \mid x \mid \text{op } \vec{e}$, where c is a constant such as a real or a string, $x \in \mathcal{N}$ is a name, op represents an operator, and $\vec{e} ::= \emptyset \mid e, \vec{e}$ is a tuple or vector of such expressions, and especially, $\vec{c}$ denotes a vector of constants. We use $|\vec{e}|$ and $\mathcal{N}(\vec{e})$ to denote the length of and the set of names in $\vec{e}$, respectively. A Boolean expression $B$ can be defined (or derived) from the strictly smaller grammar $B ::= \textbf{false} \mid e = e \mid e < e \mid B \wedge B \mid \neg B$.

### 4.2 Syntax

The syntax of **HpC**, defined below, is an extension to the classic $\pi$-calculus in [35, 36] with two constructs, highlighted in blue: (1) a general form of *guard* $[B]$, and (2) prefix $\{\vec{e}_0 \mid \dot{\vec{v}} = \vec{e}\&B, \mathcal{R}\}$ to describe the *continuous* behaviour of a physical process.

In the remainder and for clarity, we use terms P, Q, and R to represent processes and M, N, and W to denote sums $\sum_{i \in I} \pi_i.\text{P}_i$, where sub-terms guard a process P$_i$ by a prefix $\pi_i$, which can be empty (silence $\tau$), an input $x(\vec{y})$, an output $\overline{x}\langle\vec{e}\rangle$, a Boolean clause $[B]$, or a continuous $\{\vec{e}_0 \mid \dot{\vec{v}} = \vec{e}\&B, \mathcal{R}\}$. The empty sum ($I = \emptyset$) stands for *inaction* and is noted as process 0. The restriction of a name $x$ to the scope of P is noted $(\nu x)$P, its replication noted !P and its parallel composition to P′ noted P||P′.

The term $\{\vec{e}_0 \mid \dot{\vec{v}} = \vec{e}\&B, \mathcal{R}\}(\vec{y}).\mathsf{P}$ denotes a continuous process. From left to right, it consists of a vector of expressions $\vec{e}_0$ representing the initial state of the ODE's variables $\vec{v}$ and $\vec{e}$ its *vector field* which we require to be locally Lipschitz-continuous (as formally defined in Theorem 6.7). The Boolean condition $B$ defines the domain constraints (range or boundary) of the continuous variable. Once $B$ is violated by the state of the ODE, e.g. $\vec{c}$, then its evolution stops immediately and its continuation $(\vec{y}).\mathsf{P}$ is applied to the final state $\vec{c}$, reducing to $\mathsf{P}\{\vec{c}/\vec{y}\}$. Finally, the set $\mathcal{R} \subseteq \{v, \overline{v} \mid v \in \mathcal{N}(\vec{v})\}$ is called the *ready set* or *interface* of the ODE. It declares the channel names which can be sensed or actuated by discrete processes. For brevity, we omit $B$ if it is **true** and omit $\mathcal{R}$ if it is $\emptyset$.

*Definition 4.1 (Syntax).*

$$
\begin{array}{rclll}
\pi & ::= & \tau & \text{(Silent)} \\
    & \mid & x(\vec{y}) & \text{(Input)} \\
    & \mid & \overline{x}\langle \vec{e} \rangle & \text{(Output)} \\
    & \mid & [B] & \text{(Guard)} \\
    & \mid & \{\vec{e}_0 \mid \dot{\vec{v}} = \vec{e}\&B, \mathcal{R}\}(\vec{y}) & \text{(Continuous)}
\end{array}
\qquad
\begin{array}{rclll}
\mathsf{P} & ::= & \sum_{i \in I} \pi_i.\mathsf{P}_i & \text{(Sum)} \\
    & \mid & (vx)\mathsf{P} & \text{(Restriction)} \\
    & \mid & \mathsf{P} \| \mathsf{P}' & \text{(Parallel)} \\
    & \mid & !\mathsf{P} & \text{(Replication)}
\end{array}
$$

REMARK 2. *For a continuous action $\{\vec{e}_0 \mid \dot{\vec{v}} = \vec{e}\&B, \mathcal{R}\}$, i.e., we do not assume that $\mathcal{N}(\vec{e}) \subseteq \mathcal{N}(\vec{v})$ or $\mathcal{N}(B) \subseteq \mathcal{N}(\vec{v})$ (as opposed to autonomous systems, whose continuous variables are local). For instance, $\{0 \mid \dot{v} = w \& u < 1, \{v, \overline{v}\}\}$ allows to reference external physical variables $w$ and $u$.*

### 4.3 Scope of Names

We write $\mathcal{B}(\mathsf{P})$ for the names bound in $\mathsf{P}$ and conversely, $\mathcal{F}(\mathsf{P})$ for the *free* names of $\mathsf{P}$. For example, $\mathcal{F}(x(\vec{y}).\mathsf{P}) = \{x\} \cup \mathcal{F}(\mathsf{P}) - \mathcal{N}(\vec{y})$ and $\mathcal{B}(x(\vec{y}).\mathsf{P}) = \mathcal{B}(\mathsf{P}) \cup \mathcal{N}(\vec{y})$. We adopt Barendregt's Convention that the free and bound names of each process are disjoint: $\mathcal{F}(\mathsf{P}) \cap \mathcal{B}(\mathsf{P}) = \emptyset$ for all $\mathsf{P}$ and hence $\mathcal{N}(\mathsf{P}) = \mathcal{F}(\mathsf{P}) \uplus \mathcal{B}(\mathsf{P})$.

### 4.4 Structural Congruence

Structural congruence $\equiv$ is defined as the smallest congruence satisfying the following laws:

- If $\mathsf{P}$ is an $\alpha$-conversion of $\mathsf{Q}$, then $\mathsf{P} \equiv \mathsf{Q}$.
- Commutativity: $\mathsf{P} \| \mathsf{Q} \equiv \mathsf{Q} \| \mathsf{P}$ and $\mathsf{M} + \mathsf{N} \equiv \mathsf{N} + \mathsf{M}$.
- Associativity: $(\mathsf{P} \| \mathsf{Q}) \| \mathsf{R} \equiv \mathsf{P} \| (\mathsf{Q} \| \mathsf{R})$ and $(\mathsf{M} + \mathsf{N}) + \mathsf{W} \equiv \mathsf{M} + (\mathsf{N} + \mathsf{W})$.

Equivalences of $\mathsf{P} \| \mathbf{0} \equiv \mathsf{P}$, $\mathsf{M} + \mathbf{0} \equiv \mathsf{M}$, $(vx)\mathbf{0} \equiv \mathbf{0}$, $(vx)(\mathsf{P} \| \mathsf{Q}) \equiv \mathsf{P} \| (vx)\mathsf{Q}$ if $x \notin \mathcal{F}(\mathsf{P})$, and so on are proved with the strong bisimulation relation $\sim$ defined in Section 6.1.

## 5 OPERATIONAL SEMANTICS

In this section, we first define the substitutions on expressions and processes in Section 5.1. Then, we define states and flows in Section 5.2 and 5.3 to describe the dynamics of a physical process. A state maps names to reals while a flow is a function over time describing the trajectory or evolution of a physical process. Section 5.4 specifies the operational semantics of the proposed **HpC**, where we can see how continuous variables are viewed as channels sampled by discrete processes.

### 5.1 Substitutions

A substitution $\Delta \triangleq \{e_1/x_1, \cdots, e_n/x_n\}$, also written $\{e_1, \cdots, e_n/x_1, \cdots, x_n\}$ or $\{\vec{e}/\vec{x}\}$, maps pairwise distinct names $x_i$ (i.e. $i \neq j \Rightarrow x_i \neq x_j$) of a process $\mathsf{P}$ (resp. an expression $e$) to an expression $x_i \Delta = e_i$, for all $1 \leq i \leq n$, written $\mathsf{P}\Delta$ (resp. $e\Delta$), and acts as the identity otherwise.

## 5.2 States

A state $\sigma_V : V \to \mathbb{R}$ maps the continuous variables in $V \subseteq \mathcal{V}$ to reals. We note it $\sigma$ if its domain $\mathrm{dom}(\sigma) = V$ is specified. The union of two states $\sigma \cup \sigma'$ is *defined* iff. $\sigma$ and $\sigma'$ agree on common names: for all $x \in \mathrm{dom}(\sigma) \cap \mathrm{dom}(\sigma')$, $\sigma(x) = \sigma'(x)$. Let $\vec{\omega}$ be a vector of names and constants. The evaluation of $\vec{e}$ in state $\sigma$, denoted $[\![\vec{e}]\!]_\sigma$, is defined by the following rules:

$$\frac{}{[\![c]\!]_\sigma = c} \quad \frac{x \in \mathrm{dom}(\sigma)}{[\![x]\!]_\sigma = \sigma(x)} \quad \frac{x \notin \mathrm{dom}(\sigma)}{[\![x]\!]_\sigma = x} \quad \frac{\mathsf{op}\,[\![\vec{e}]\!]_\sigma = \vec{\omega}}{[\![\mathsf{op}\,\vec{e}]\!]_\sigma = \vec{\omega}} \quad \frac{}{[\![e, \vec{e}]\!]_\sigma = [\![e]\!]_\sigma, [\![\vec{e}]\!]_\sigma}$$

For brevity, $[\![\vec{e}]\!]$ denotes $[\![\vec{e}]\!]_{\sigma_\emptyset}$, for example, $[\![1+2]\!] = 3$, $[\![x]\!] = x$, and $[\![x = x]\!] = \mathbf{true}$, while $[\![1/0]\!]$, $[\![x+1]\!]$, and $[\![x > y]\!]$ are undefined. Evaluation acts as substitution, hence $[\![x]\!]_\sigma = x$ if $x \notin \mathrm{dom}(\sigma)$.

## 5.3 Flows

A flow $\rho_V^\mathrm{d} : [0, \mathrm{d}) \to V \to \mathbb{R}$ is defined by a function over time using a set of continuous variables $V \subseteq \mathcal{V}$. $\rho_V^\mathrm{d}(t)$ designates the state of $V$ at time $t \in [0, \mathrm{d})$. For brevity, we write $\mathrm{Left}(\rho_V^\mathrm{d}) \triangleq \rho_V^\mathrm{d}(0)$ and $\mathrm{Right}^-(\rho_V^\mathrm{d}) \triangleq \lim_{t \to \mathrm{d}^-} \rho_V^\mathrm{d}(t)$ for the left and right-limit states of $\rho_V^\mathrm{d}$, respectively. The name set and the duration of $\rho_V^\mathrm{d}$ are obtained by $\mathcal{N}(\rho_V^\mathrm{d}) = V$ and $|\rho_V^\mathrm{d}| = \mathrm{d}$. Therefore, in what follows, we omit $V$ and $\mathrm{d}$ and use $\rho$ to denote a flow. We require $\rho$ to be *right continuous* and *semi-differentiable* within its domain. Additionally, the left limit at the end point of any flow should exist:

**Right Continuity** $\forall v \in \mathcal{N}(\rho) \,.\, \forall t \in [0, |\rho|) \,.\, \rho(t)(v) = \lim_{\varepsilon \to 0^+} \rho(t+\varepsilon)(v)$.
**Semi-differentiability** $\forall v \in \mathcal{N}(\rho) \,.\, \forall t \in (0, |\rho|) \,.\, \exists c_1, c_2 \in \mathbb{R} \,.\, \dot\rho(t^-)(v) = c_1 \land \dot\rho(t^+)(v) = c_2$,
  where $\dot\rho(t^\pm)(v) \triangleq \lim_{\varepsilon \to 0^\pm} (\rho(t+\varepsilon)(v) - \rho(t)(v))/\varepsilon$ for $\pm \in \{+, -\}$.
**Left Limit** $\forall v \in \mathcal{N}(\rho) \,.\, \exists c \in \mathbb{R} \,.\, \mathrm{Right}^-(\rho)(v) = c$.

Two flows $\rho$ and $\rho'$ can be concatenated, denoted $\rho \frown \rho'$, if the right limit of $\rho$ and the initial state of $\rho'$ coincide, i.e. $\mathrm{Right}^-(\rho) = \mathrm{Left}(\rho')$. Formally, $(\rho \frown \rho')(t) = \rho(t)$ if $0 \le t < |\rho|$ and $\rho'(t - |\rho|)$ if $|\rho| \le t < |\rho| + |\rho'|$. The union and intersection of two flows are defined if they have the same duration and agree on the common names. We define $\rho \uplus \rho' = \rho \cup \rho'$ if $\mathcal{N}(\rho) \cap \mathcal{N}(\rho') = \emptyset$. In addition, $\rho - \rho' = \rho''$ if $\rho = \rho'' \uplus (\rho \cap \rho')$.

A flow $\rho$ can be refined as a pair $\mathcal{A} \cdot \mathcal{G}$, logically equivalent to $\mathcal{A} \uplus \mathcal{G}$, where $\mathcal{G}$ is the flow of the continuous variables defined in the process under consideration and $\mathcal{A}$ is the flow of the (undefined) continuous variables it uses. In other words, $\mathcal{A}$ is the trajectory of continuous variables defined in the environment with which the process interacts. In this sense, $\mathcal{A}$ and $\mathcal{G}$ may be called the *assumption* and *guarantee*, respectively, in terms borrowed from contract theory [9]. A flow $\mathcal{A} \cdot \mathcal{G}$ is said to be *closed* if $\mathcal{N}(\mathcal{A}) = \emptyset$, i.e., it has no assumption on the environment. In control theory, it is hence called an autonomous system (the actual implementation of a physical process). The composition of flows can algebraically be defined as the composition of contracts:

$$(\mathcal{A} \cdot \mathcal{G}) \otimes (\mathcal{A}' \cdot \mathcal{G}') \quad \triangleq \quad \big((\mathcal{A} - \mathcal{G}') \cup (\mathcal{A}' - \mathcal{G})\big) \cdot (\mathcal{G} \cup \mathcal{G}')$$

In the remainder, we use $\rho$ and $\mathcal{A} \cdot \mathcal{G}$ to denote a flow, interchangeably. We say a set of ODEs $\dot{\vec{v}} = \vec{e}$ evolves along a flow $\mathcal{A} \cdot \mathcal{G}$, denoted $[\![\dot{\vec{v}} = \vec{e}]\!]_{\mathcal{A} \cdot \mathcal{G}}$, iff (1) $\mathcal{N}(\mathcal{G}) = \mathcal{N}(\vec{v})$ and $\mathcal{N}(\mathcal{A}) = \mathcal{N}(\vec{e}) - \mathcal{N}(\vec{v})$, and (2) $\forall t \in (0, |\mathcal{A} \cdot \mathcal{G}|) \,.\, \dot{\mathcal{G}}(t)(\vec{v}) = [\![\vec{e}]\!]_{(\mathcal{A} \cdot \mathcal{G})(t)}$. Similarly, we say a Boolean expression $B$ keeps holding along a flow $\rho$, denoted $[\![B]\!]_\rho$, iff $\forall t \in [0, |\rho|) \,.\, [\![B]\!]_{\rho(t)} = \mathbf{true}$.

## 5.4 A Structural Operational Semantics

The operational semantics is defined by a proof system that consists of two rule-sets:

(1) A set of discrete transition rules, for example, $\mathsf{P} \xrightarrow{\tau} \mathsf{Q}$.
(2) A set of, so called, continuous evolution rules, for example, $\mathsf{P} \xrightarrow{\mathrm{d}} \mathsf{Q}$ where $\mathrm{d} > 0$.

While the first one defines transitions that can conceptually be perceived as taking "discrete" or no time to perform an action, like a conventional labelled transition system, the second rule-set does, on the contrary, define the continuous trajectories of an hybrid system over time.

### 5.4.1 Abstractions and Concretions.
To express the discrete transition relation correctly, we introduce the notions of *abstractions* and *concretions* due to [36].

*Definition 5.1 (Abstraction and concretion).* $(\vec{x}).P$ is an *abstraction* of process P. Two abstractions are structurally congruent ($\equiv$) if, up to $\alpha$-conversion, their bound names $\vec{x}$ are identical and their processes are structurally congruent. Conversely, $(\nu\vec{y})\langle\vec{e}\rangle.P$ is a *concretion* of process P with $\mathcal{N}(\vec{y}) \subseteq \mathcal{N}(\vec{e})$. Two concretions are structurally congruent if, up to alpha-conversion and re-ordering of restricted names, their prefixes $(\nu\vec{y})\langle\vec{e}\rangle$ are identical and their processes are structurally congruent.

We highlight *abstractions* as F or G and *concretions* as C or D, as they have a specific usage. An *agent* A or B is either an abstraction or a concretion. An ordinary process P is a special kind of agent: it is both an abstraction ().P and a concretion $\langle\rangle$.P.

*Definition 5.2 (Ions).* Abstraction and a concretion can be regarded as positive and negative *ions*. Assuming $y \notin \mathcal{N}(\vec{x})$ and $\mathcal{N}(\vec{x}) \cap \mathcal{F}(Q) = \emptyset$, then:

$$(\nu y)(\vec{x}).P \triangleq (\vec{x}).(\nu y)P \text{ and } ((\vec{x}).P) \| Q \triangleq (\vec{x}).(P\|Q)$$

$$((\nu\vec{x})\langle\vec{e}\rangle.P) \| Q \triangleq (\nu\vec{x})\langle\vec{e}\rangle.(P\|Q) \text{ and } (\nu y)((\nu\vec{x})\langle\vec{e}\rangle.P) \triangleq \begin{cases} (\nu y, \vec{x})\langle\vec{e}\rangle.P & \text{if } y \in \mathcal{N}(\vec{e}) \\ (\nu\vec{x})\langle\vec{e}\rangle.(\nu y)P & \text{otherwise} \end{cases}$$

*Definition 5.3 (Application).* The application of an abstraction and a concretion is defined by $(\vec{x}).P@(\nu\vec{y})\langle\vec{e}\rangle.Q \triangleq (\nu\vec{y})(P\{\vec{e}/\vec{x}\}\|Q)$, assuming that $\mathcal{N}(\vec{y}) \cap \mathcal{F}(P) \subseteq \mathcal{N}(\vec{x})$. In particular, for an abstraction $F = (\vec{x}).P$, we write $F\langle\vec{e}\rangle$ for $F@\langle\vec{e}\rangle$ to mean $P\{\vec{e}/\vec{x}\}$.

### 5.4.2 Small-step Operational Semantics.
The transition of a process P to an agent A performing the action $\alpha$ is defined by $P \xrightarrow{\alpha} A$. The action $\alpha$ can be discrete or continuous:

(1) A discrete action $\lambda$, which can either be silent ($\tau$), an input $x$, or an output $\overline{x}$;
(2) A continuous action $\langle\rho, \mathcal{R}\rangle$, which describes the continuous trajectory defined by a flow $\rho$ and a ready set of channels $\mathcal{R}$, pending communication during the evolution.

The operational semantics is defined by the rule-sets of Table 1, for discrete transitions, and of Table 2, for continuous evolution. We first define the structural equivalence rule. It stipulates that, if two processes are structurally congruent, then they have the same behaviour. Rule [Eqv] provides alpha-equivalence, commutativity, associativity to both transition systems: it is labelled by $\alpha$.

$$\frac{P \equiv Q \quad Q \xrightarrow{\alpha} B \quad B \equiv A}{P \xrightarrow{\alpha} A}[\text{Eqv}]$$

Rule [$\lambda$-Act]. Top-left of Table 1, a sum is represented as $\lambda A + M$, where $\lambda$ is a discrete action (silence $\tau$, input $x$, or output $\overline{x}$) and A is an agent (abstraction, concretion, or process). The agent A is triggered if $\lambda$ is executed.

Rule [Pass]. If the Boolean expression $B$ evaluates to **true**, the process $[B].P+M$ silently transitions to P. Guards allow to define **if** $B$ **then** P **else** Q by $[B].P + [\neg B].Q$

Rule [Sense]. A process of continuous prefix can be selected for sensing to allow an ODE be interrupted and sensed by the environment. Assume that $v$ is the $i$-th element of $\vec{v}$ in the ODE: $v = \vec{v}(i)$. Then, the output action $\overline{v}$ in the interface $\mathcal{R}$ indicates that the current value of $v$ $e = \vec{e}_0(i)$ can be sent out by $\overline{v}$. Thus, the continuous process transitions to a concretion whose "charge" is $\langle e \rangle$ by the output action $\overline{v} \in \mathcal{R}$. Notice that sensing does not stop the ODE. After merging the "charge" $\langle e \rangle$ with the environment, the ODE continues to evolve from its current state $\vec{e}_0$.

Rule [Actuate]. Actuation is symmetric to sensing: a process of continuous prefix can be interrupted for actuation. The presence of an input action $v$ in the interface $\mathcal{R}$ indicates that the current value of $v$, i.e., $\vec{e}_0(i)$, can be modified to a new value, denoted by a fresh variable $x$. When that happens, the $i$-th element of $\vec{e}_0$ is replaced with $x$, represented by $\vec{e}_1 = \vec{e}_0[i \mapsto x]$ (to mean that $\vec{e}_1(i) = x$ and $\vec{e}_1(j) = \vec{e}_0(j)$ for all $j \neq i$). In the conclusion of the rule, the continuous process transitions to an abstraction whose "charge" is $(x)$ by an input action $v \in \mathcal{R}$. Similarly, actuation does not stop an ODE, i.e., once an actual value substitutes the "charge" $(x)$ by composition, the resumes its evolution with its updated state.

Table 1. Discrete transitions

$$\frac{}{\lambda A + M \xrightarrow{\lambda} A}[\lambda\text{-Act}] \qquad \frac{[\![B]\!] = \mathbf{true}}{[B].P + M \xrightarrow{\tau} P}[\text{Pass}]$$

$$\frac{\bar{v} \in \mathcal{R} \quad v = \vec{v}(i) \quad \vec{e}_0(i) = e}{\{\vec{e}_0 \mid \dot{\vec{v}} = \vec{e}\&B, \mathcal{R}\}(\vec{y}).P + M \xrightarrow{\bar{v}} \langle e \rangle.\{\vec{e}_0 \mid \dot{\vec{v}} = \vec{e}\&B, \mathcal{R}\}(\vec{y}).P}[\text{Sense}]$$

$$\frac{v \in \mathcal{R} \quad v = \vec{v}(i) \quad \vec{e}_1 = \vec{e}_0[i \mapsto x] \quad x \notin \mathcal{N}(\vec{e}_0, \vec{v}, \vec{e}, B, \vec{z}) \cup \mathcal{N}(P)}{\{\vec{e}_0 \mid \dot{\vec{v}} = \vec{e}\&B, \mathcal{R}\}(\vec{y}).P + M \xrightarrow{v} (x).\{\vec{e}_1 \mid \dot{\vec{v}} = \vec{e}\&B, \mathcal{R}\}(\vec{y}).P}[\text{Actuate}]$$

$$\frac{P \xrightarrow{\lambda} A}{P\|Q \xrightarrow{\lambda} A\|Q}[\text{Par}] \quad \frac{P \xrightarrow{x} F \quad Q \xrightarrow{\bar{x}} C}{P\|Q \xrightarrow{\tau} F@C}[\text{Sync}] \quad \frac{P \xrightarrow{\lambda} A \quad \lambda \notin \{x, \bar{x}\}}{(\nu x)P \xrightarrow{\lambda} (\nu x)A}[\text{Res}] \quad \frac{P\|!P \xrightarrow{\lambda} A}{!P \xrightarrow{\lambda} A}[\text{Rep}]$$

Rule [Sync]. As for the rule of sensing and actuation, an input action $x$ produces an abstraction $F$ which synchronises with the concretion $C$ of the matching output action $\bar{x}$. It results in a silent action $\tau$ and the application $F@C$.

Rule [Par]. Parallel processes $P\|Q$ may also evolve independently w.r.t. a discrete action $\lambda$.

Rule [Res]. Restriction $(\nu x)P$ behaves as $P$ if its discrete action $\lambda$ does not concern the name $x$, i.e., $\lambda \notin \{x, \bar{x}\}$. The rule stipulates that references to $x$ cannot escape the scope of $(\nu x)(\cdot)$.

Rule [Rep]. The meaning of replication $!P$ reflects the intuition of forking an unlimited number of copies of $P$ in parallel. Replication can model recursion as $\mu x(\vec{y}).P@\langle \vec{e} \rangle \triangleq (\nu x)(\bar{x}\langle \vec{e}\rangle\|!x(\vec{y}).P)$. For convenience, we write $\mu x.P$ for $\mu x().P@\langle\rangle$ as, e.g., in the recursion Ground in Section 3.1.

The rule-sets in Table 2 define the continuous behaviours of all (hybrid) processes.

Rule [Run]. A run describes the evolution of a continuous prefix $\{\vec{e}_0 \mid \dot{\vec{v}} = \vec{e}\&B, \mathcal{R}\}$. It starts from the initial state represented by a vector of expressions $\vec{e}_0$ and then evolves according to the (set of) ODE(s) $\dot{\vec{v}} = \vec{e}$. Thus, $\vec{e}_0$ should be evaluated with respect to the initial state of $\mathcal{G}$, i.e., $\text{Left}(\mathcal{G})(\vec{v}) = [\![\vec{e}_0]\!]$. Then, the derivative $\dot{\vec{v}} = \vec{e}$ must evaluate to **true** along the entire trajectory $\mathcal{A} \cdot \mathcal{G}$, i.e., $[\![\dot{\vec{v}} = \vec{e}]\!]_{\mathcal{A} \cdot \mathcal{G}}$. The state of the ODE evolves within the boundary defined by $B$, i.e., $[\![B]\!]_{\mathcal{A} \cdot \mathcal{G}}$ while the channels in the ready set $\mathcal{R}$ listen to the environment. The final state of $\vec{v}$ of the evolution is defined by the limit $\text{Right}^-(\mathcal{G})(\vec{v}) = \vec{c}$. However, the premise $[\![B]\!]_{\text{Right}^-(\mathcal{A} \cdot \mathcal{G})}$ should still hold when that happens, allowing the process to resume from state $\vec{c}$ and, e.g., an interruption using rules [Sense] or [Actuate] of Table 1.

Rule [Stop]. Continuous evolution stops once it reaches the boundary of $B$. Starting from the initial condition $[\![\vec{e}_0]\!]$, the ODE evolves along the trajectory $\mathcal{A} \cdot \mathcal{G}$ until it reaches its boundary, i.e., $\neg[\![B]\!]_{\text{Right}^-(\mathcal{A} \cdot \mathcal{G})}$, unlike rule [Run]. Once the evolution stops, the current state $\vec{c} = \text{Right}^-(\mathcal{G})(\vec{v})$ of the ODE is transferred to its continuation, $(\vec{y}).P$, i.e., $(\vec{y}).P@\langle \vec{c} \rangle \equiv P\{\vec{c}/\vec{y}\}$.

Table 2. Continuous evolution

$$\frac{\text{Left}(\mathcal{G})(\vec{v}) = [\![\vec{e}_0]\!] \quad \text{Right}^-(\mathcal{G})(\vec{v}) = \vec{c} \quad [\![\dot{\vec{v}} = \vec{e}]\!]_{\mathcal{A}\cdot\mathcal{G}} \quad [\![B]\!]_{\mathcal{A}\cdot\mathcal{G}} \quad [\![B]\!]_{\text{Right}^-(\mathcal{A}\cdot\mathcal{G})}}{\{\vec{e}_0 \mid \dot{\vec{v}} = \vec{e}\&B, \mathcal{R}\}(\vec{y}).\mathsf{P} + \mathsf{M} \xrightarrow{\langle \mathcal{A}\cdot\mathcal{G},\mathcal{R}\rangle} \{\vec{c} \mid \dot{\vec{v}} = \vec{e}\&B, \mathcal{R}\}(\vec{y}).\mathsf{P}} \text{[Run]}$$

$$\frac{\text{Left}(\mathcal{G})(\vec{v}) = [\![\vec{e}_0]\!] \quad \text{Right}^-(\mathcal{G})(\vec{v}) = \vec{c} \quad [\![\dot{\vec{v}} = \vec{e}]\!]_{\mathcal{A}\cdot\mathcal{G}} \quad [\![B]\!]_{\mathcal{A}\cdot\mathcal{G}} \quad \neg[\![B]\!]_{\text{Right}^-(\mathcal{A}\cdot\mathcal{G})}}{\{\vec{e}_0 \mid \dot{\vec{v}} = \vec{e}\&B, \mathcal{R}\}(\vec{y}).\mathsf{P} + \mathsf{M} \xrightarrow{\langle \mathcal{A}\cdot\mathcal{G},\mathcal{R}\rangle} \mathsf{P}\{\vec{c}/\vec{y}\}} \text{[Stop]}$$

$$\frac{\forall i \in I \,.\, \lambda_i \neq \tau}{\sum_{i\in I} \lambda_i \mathsf{A}_i \xrightarrow{\langle \rho_\emptyset^d, \{\lambda_i\}_{i\in I}\rangle} \sum_{i\in I} \lambda_i \mathsf{A}_i} \text{[Wait]} \qquad \frac{\mathsf{P} \xrightarrow{\langle \rho,\mathcal{R}\rangle} \mathsf{P}' \quad \mathsf{Q} \xrightarrow{\langle \rho',\mathcal{R}'\rangle} \mathsf{Q}' \quad \overline{\mathcal{R}} \cap \mathcal{R}' = \emptyset}{\mathsf{P} \| \mathsf{Q} \xrightarrow{\langle \rho\otimes\rho',\mathcal{R}\cup\mathcal{R}'\rangle} \mathsf{P}' \| \mathsf{Q}'} \text{[Par'}]$$

$$\frac{\mathsf{P} \xrightarrow{\langle \mathcal{A}\cdot\mathcal{G},\mathcal{R}\rangle} \mathsf{P}' \quad x \notin \mathcal{N}(\mathcal{A})}{(\nu x)\mathsf{P} \xrightarrow{\langle \mathcal{A}\cdot\mathcal{G},\mathcal{R}\rangle\backslash x} (\nu x)\mathsf{P}'} \text{[Res'}] \qquad \frac{\mathsf{P} \xrightarrow{\langle \rho,\mathcal{R}\rangle} \mathsf{P}' \quad \overline{\mathcal{R}} \cap \mathcal{R} = \emptyset}{!\mathsf{P} \xrightarrow{\langle \rho,\mathcal{R}\rangle} !\mathsf{P}'} \text{[Rep'}]$$

*Example 5.4.* Consider $\{1 \mid \dot{v} = v \,\&\, v < 5\}(y).\mathsf{P}$, where the solution of $\{\dot{v} = v\}$ with initial condition $v(0) = 1$ is $v(t) = \exp(t)$ for $t > 0$. It may evolve for $\ln(5)$ time units and then stop to execute the continuation by rule [Stop]:

$$\frac{\text{Left}(\mathcal{G})(v) = [\![1]\!] \quad \text{Right}^-(\mathcal{G})(v) = 5 \quad [\![\dot{v} = v]\!]_{\mathcal{A}\cdot\mathcal{G}} \quad [\![v < 5]\!]_{\mathcal{A}\cdot\mathcal{G}} \quad \neg[\![v < 5]\!]_{\text{Right}^-(\mathcal{A}\cdot\mathcal{G})}}{\{1 \mid \dot{v} = v \,\&\, v < 5\}(y).\mathsf{P} \xrightarrow{\langle \mathcal{A}\cdot\mathcal{G},\emptyset\rangle} \mathsf{P}\{5/y\}}$$

where $\mathcal{A} \cdot \mathcal{G}$ is the flow with $|\mathcal{A} \cdot \mathcal{G}| = \ln(5)$, $\mathcal{N}(\mathcal{A}) = \emptyset$, $\mathcal{N}(\mathcal{G}) = \{v\}$, and $\mathcal{G}(t)(v) = \exp(t)$. Also, the process may evolve for mere 1 time units by rule [Run], i.e., $|\mathcal{A} \cdot \mathcal{G}| = 1$, $\mathcal{N}(\mathcal{A}) = \emptyset$, $\mathcal{N}(\mathcal{G}) = \{v\}$, and $\mathcal{G}(t)(v) = \exp(t)$. The target process may then evolve for $\ln(5) - 1$ time units to reach $\mathsf{P}\{5/y\}$, i.e., the flow from $\{1 \mid \dot{v} = v \,\&\, v < 5\}(y).\mathsf{P}$ to $\mathsf{P}\{5/y\}$ is chopped into two segments.

$$\frac{\text{Left}(\mathcal{G})(v) = [\![1]\!] \quad \text{Right}^-(\mathcal{G})(v) = \exp(1) \quad [\![\dot{v} = v]\!]_{\mathcal{A}\cdot\mathcal{G}} \quad [\![v < 5]\!]_{\mathcal{A}\cdot\mathcal{G}} \quad [\![v < 5]\!]_{\text{Right}^-(\mathcal{A}\cdot\mathcal{G})}}{\{1 \mid \dot{v} = v \,\&\, v < 5\}(y).\mathsf{P} \xrightarrow{\langle \mathcal{A}\cdot\mathcal{G},\emptyset\rangle} \{\exp(1) \mid \dot{v} = v \,\&\, v < 5\}(y).\mathsf{P}}$$

Rule [Wait]. In the continuous semantics, a sum $\sum_{i\in I} \lambda_i \mathsf{A}_i$ holding discrete input and output prefixes (i.e. $\lambda_i \neq \tau$) is regarded as choice waiting for an interruption to happen. To that effect, its prefixes are added to the ready set as $\{\lambda_i\}_{i\in I}$ and its flow is empty ($\rho_\emptyset^d$).

Rule [Par']. Two processes evolving in parallel, $\mathsf{P} \| \mathsf{Q}$, can synchronise over the same period of time by jointly orchestrating the "contractual" composition $\rho \otimes \rho'$, if one is not waiting for the other to communicate, i.e., $\overline{\mathcal{R}} \cap \mathcal{R}' = \emptyset$, where $\overline{\mathcal{R}} \triangleq \{\overline{x}, y \mid x, \overline{y} \in \mathcal{R}\}$ is the dual of $\mathcal{R}$. Conversely, if $\overline{\mathcal{R}} \cap \mathcal{R}' \neq \emptyset$, both parties are waiting for a discrete communication to occur with a matching channel. This must instead be done immediately, by using rules [Sense], [Actuate] or [Sync], instead of rule [Par'], which does not apply in that case.

*Example 5.5.* Consider the parallel composition of $x(y).\mathsf{P}$ and $\overline{x}\langle 1\rangle.\mathsf{Q}$, it can be resolved by using rule [Sync] in Table 1, instead of [Par'] in Table 2, because the ready sets match: $\overline{\{x\}} \cap \{\overline{x}\} = \{\overline{x}\}$.

Rule [Res']. Similarly, if $\mathsf{P}$ can perform the trajectory $\mathcal{A} \cdot \mathcal{G}$ if the scope of $x$ is restricted to $\mathsf{P}$ (it is not an assumption $\mathcal{A}$), then $(\nu x)\mathsf{P}$ can perform the same trajectory while keeping the name $x$ private to $\mathsf{P}$, represented by $\langle \mathcal{A} \cdot \mathcal{G}, \mathcal{R}\rangle \backslash x$, which is equivalent to $\langle \mathcal{A} \cdot \mathcal{G} \backslash x, \mathcal{R} - \{x, \overline{x}\}\rangle$, where $\mathcal{G} \backslash x = \mathcal{G}$ if $x \notin \mathcal{N}(\mathcal{G})$ and $\mathcal{G}'$ if $\mathcal{G} = \mathcal{G}' \uplus \mathcal{G}_x$ with $\mathcal{N}(\mathcal{G}_x) = \{x\}$.

Rule [Rep']. A continuous transition from P to P' implies one from !P to !P'. Intuitively, it is a special case of rule [Par'], since !P can actually be seen as copies of P executing in parallel.

REMARK 3. *A continuous action of the form $\langle \rho_\emptyset^d, \emptyset \rangle$ denotes an evolution that lasts for d time, whose trajectory cannot be observed (it has no observable variable in lexical scope) and whose ready set is empty. It hence can be treated by its side-effect: a delay of time, and simplified as d, the duration of the delay. For example, the transition from* wait(d) *in Section 3 may adopt this simplification.*

## 6 BISIMULATION AND CONGRUENCE

The **HpC** is a conservative extension of the classical $\pi$-calculus. As we have seen, the extension is operationally "minimal", and is yet able to describe mobility and physics while allowing to lift existing theoretical results (e.g. bisimulation) of the classic $\pi$-calculus [36] to the context of that extension. We first define strong and weak bisimulations in the **HpC**, extended to hybrid systems, and prove some of its desired properties. Then, we define the notion of approximate bisimulation, which can be treated as the generalised form of weak bisimulation. It is worth mentioning that approximate bisimulation is, to date, only known for first-order calculi (i.e. without process dynamicity or channel mobility).

### 6.1 Strong Bisimulation and Congruence

*Definition 6.1 (Strong Bisimulation).* Let F and G denote abstractions and C and D concretions, a binary relation $\mathcal{S}$ over the processes is a strong simulation if, for each P$\mathcal{S}$Q we have

- P $\xrightarrow{\tau}$ P' implies $\exists$Q' s.t. Q $\xrightarrow{\tau}$ Q' and P'$\mathcal{S}$Q';
- P $\xrightarrow{x}$ F implies $\exists$G s.t. Q $\xrightarrow{x}$ G and F$\mathcal{S}$G;
- P $\xrightarrow{\overline{x}}$ C implies $\exists$D s.t. Q $\xrightarrow{\overline{x}}$ D and C$\mathcal{S}$D;
- P $\xrightarrow{\langle \rho_1 \frown \cdots \frown \rho_n, \mathcal{R} \rangle}$ P' implies $\exists$Q' s.t. Q $\xrightarrow{\langle \rho_1, \mathcal{R} \rangle} \cdots \xrightarrow{\langle \rho_n, \mathcal{R} \rangle}$ Q' and P'$\mathcal{S}$Q'.

The $\mathcal{S}$-relations between abstractions and concretions are defined below:

- F$\mathcal{S}$G means that F$\langle \vec{e} \rangle \mathcal{S}$G$\langle \vec{e} \rangle$ for all $\vec{e}$;
- C$\mathcal{S}$D means that C $\equiv (\nu \vec{y}) \langle \vec{e} \rangle$.P and D $\equiv (\nu \vec{y}) \langle \vec{e} \rangle$.Q s.t. P$\mathcal{S}$Q.

If both $\mathcal{S}$ and its inverse are strong simulations then $\mathcal{S}$ is a strong bisimulation. Two agents A and B are strongly equivalent, written A $\sim$ B, if (A, B) belongs to a strong bisimulation.

The following two theorems prove that $\sim$ for the **HpC** is (1) an equivalence relation, i.e., P $\sim$ P (reflexivity), P $\sim$ Q implies Q $\sim$ P (symmetry), and P $\sim$ Q and Q $\sim$ R imply P $\sim$ R (transitivity); (2) a strong bisimulation, i.e., relation $\mathcal{S} = \{(P, Q) \mid P \sim Q\}$ is a strong bisimulation, where P and Q are **HpC** processes; (3) a congruence, i.e., P $\sim$ Q implies $\mathbb{C}[P] \sim \mathbb{C}[Q]$ for any context $\mathbb{C}$.

THEOREM 6.2 (EQUIVALENCE AND STRONG BISIMULATION).

*(1) $\sim$ is an equivalence relation;*
*(2) $\sim$ is itself a strong bisimulation.*

PROOF. (1) For reflexivity, it is enough to show that the identity relation over the set of agents $\mathcal{A}$, i.e., the relation $\{(A, A) \mid A \in \mathcal{A}\}$ is a strong bisimulation. For symmetry, we have to show that if $\mathcal{S}$ is a strong bisimulation then so is its converse $\mathcal{S}^{-1}$, and this is obvious from the definition. For transitivity, we must show that if $\mathcal{S}$ and $\mathcal{S}'$ are strong bisimulations, then so is their composition

$$\mathcal{S}\mathcal{S}' \triangleq \{(A, B) \mid \exists X . A\mathcal{S}X \wedge X\mathcal{S}'B\}$$

It is enough to show that this is a strong simulation. Let $(A, B) \in \mathcal{S}\mathcal{S}'$. Then,

**Case 1** A and B are abstractions. According to Definition 6.1, $A\langle\vec{e}\rangle\mathcal{S}X\langle\vec{e}\rangle$ and $X\langle\vec{e}\rangle\mathcal{S}'B\langle\vec{e}\rangle$ for all $\vec{e}$, implying $A\langle\vec{e}\rangle\mathcal{SS}'B\langle\vec{e}\rangle$ for all $\vec{e}$.

**Case 2** A and B are concretions. Let $A \equiv (\nu\vec{y})\langle\vec{e}\rangle.P$, $B \equiv (\nu\vec{y})\langle\vec{e}\rangle.Q$, and $X \equiv (\nu\vec{y})\langle\vec{e}\rangle.R$. Then, according to Definition 6.1, $P\mathcal{S}R$ and $R\mathcal{S}'Q$, implying $P\mathcal{SS}'Q$.

**Case 3** A and B are processes. Let $A \xrightarrow{\alpha} A'$. Then,

- $\alpha = \lambda$ is a discrete action. By Definition 6.1, $\exists X'$ such that $X \xrightarrow{\lambda} X'$ and $A'\mathcal{S}X'$, and hence $\exists B'$ such that $B \xrightarrow{\lambda} B'$ and $X'\mathcal{S}'B'$. Thus, $A \xrightarrow{\lambda} A'$ implies $B \xrightarrow{\lambda} B'$ and $A'\mathcal{SS}'B'$.

- $\alpha = \langle\rho, \mathcal{R}\rangle$ is a continuous action. By Definition 6.1, $\exists X'$ s.t. $X \xrightarrow{\langle\rho_1,\mathcal{R}\rangle} \cdots \xrightarrow{\langle\rho_n,\mathcal{R}\rangle} X'$ and $A'\mathcal{S}X'$, where $\rho = \rho_1 \frown \cdots \frown \rho_n$, and hence $\exists B'$ s.t.

$$B \xrightarrow{\langle\rho_1^{(1)},\mathcal{R}\rangle} \cdots \xrightarrow{\langle\rho_1^{(m_1)},\mathcal{R}\rangle} \cdots \xrightarrow{\langle\rho_n^{(1)},\mathcal{R}\rangle} \cdots \xrightarrow{\langle\rho_n^{(m_n)},\mathcal{R}\rangle} B'$$

and $X'\mathcal{S}'B'$, where $\rho_i = \rho_i^{(1)} \frown \cdots \frown \rho_i^{(m_i)}$ for $1 \leq i \leq n$. Thus,

$$A \xrightarrow{\langle\rho_1^{(1)} \frown \cdots \frown \rho_1^{(m_1)} \frown \cdots \frown \rho_n^{(1)} \frown \cdots \frown \rho_n^{(m_n)},\mathcal{R}\rangle} A'$$

implies $B \xrightarrow{\langle\rho_1^{(1)},\mathcal{R}\rangle} \cdots \xrightarrow{\langle\rho_1^{(m_1)},\mathcal{R}\rangle} \cdots \xrightarrow{\langle\rho_n^{(1)},\mathcal{R}\rangle} \cdots \xrightarrow{\langle\rho_n^{(m_n)},\mathcal{R}\rangle} B'$ and $A'\mathcal{SS}'B'$.

In summary, the strong simulation condition for $\mathcal{SS}'$ is established.

(2) Let $A \sim B$, where A and B are agents. Then by definition $A\mathcal{S}B$ for some strong bisimulation $\mathcal{S}$.

**Case 1** If $A \xrightarrow{\lambda} A'$, then $\exists B'$ such that $B \xrightarrow{\lambda} B'$ and $A'\mathcal{S}B'$;

**Case 2** If $A \xrightarrow{\langle\rho_1 \frown \cdots \frown \rho_n, \mathcal{R}\rangle} A'$, then $\exists B'$ such that $B \xrightarrow{\langle\rho_1,\mathcal{R}\rangle} \cdots \xrightarrow{\langle\rho_n,\mathcal{R}\rangle} B'$ and $A'\mathcal{S}B'$.

Hence also $A' \sim B'$. Thus, $\sim$ is a strong simulation, and by symmetry so is its converse. □

THEOREM 6.3 (CONGRUENCE).
- If $A \sim B$ then $\lambda A + M \sim \lambda B + M$.
- If $P \sim Q$ then (1) $[B].P + M \sim [B].Q + M$, (2) $(\nu x)P \sim (\nu x)Q$, (3) $P\|R \sim Q\|R$, (4) $!P \sim !Q$, (5) $(\nu\vec{x})\langle\vec{e}\rangle.P \sim (\nu\vec{x})\langle\vec{e}\rangle.Q$, and (6) $\{\vec{e}_0 \mid \vec{v} = \vec{e}\&B, \mathcal{R}\}(\vec{y}).P + M \sim \{\vec{e}_0 \mid \vec{v} = \vec{e}\&B, \mathcal{R}\}(\vec{y}).Q + M$.
- If $P\{\vec{e}/\vec{x}\} \sim Q\{\vec{e}/\vec{x}\}$ for all $\vec{e}$, then $(\vec{x}).P \sim (\vec{x}).Q$.

PROOF. The detailed proof can be found in Appendix A. □

PROPERTY 6.1 (STRONG EQUIVALENCE).
(1) $(\nu x)0 \sim 0$, $P\|0 \sim P$, and $M + 0 \sim M$.
(2) $(\nu x)(P\|Q) \sim P\|(\nu x)Q$ if $x \notin \mathcal{F}(P)$.
(3) $(\nu x)(\nu y)P \sim (\nu y)(\nu x)P$.

PROOF. The property is proved by case analysis on the semantics rules of Tables 1 and 2 using Definition 6.1. □

The above equivalence relations usually appear as the structural congruence rules of $\pi$-calculus. Here we can prove that they are strong equivalences and hence removed from the structural congruence rules for the **HpC** (Section 4.4).

## 6.2 Weak Bisimulation

Intuitively, the difference between strong and weak bisimulation is that the latter disregards $\tau$-actions, as $\tau$-actions usually denote internal or invisible behaviours such as internal communications and discrete computations whose time can be ignored.

We use $\Rightarrow$ to denote finite $\tau$-actions. Concretely, $P \Rightarrow P'$ if $P \xrightarrow{\tau} \cdots \xrightarrow{\tau} P'$. Besides, $P \xrightarrow{x\vec{e}} P'$ is defined if $P \xrightarrow{x} F$ and $F\langle \vec{e} \rangle = P'$. Then, we can define the weak bisimulation as follows:

*Definition 6.4 (Weak Bisimulation).* A binary relation $\mathcal{S}$ over the processes is a weak simulation if, for each $P\mathcal{S}Q$ we have

- $P \xrightarrow{\tau} P'$ implies $\exists Q'$ s.t. $Q \Rightarrow Q'$ and $P'\mathcal{S}Q'$;
- $P \xrightarrow{x\vec{e}} P'$ implies $\exists Q'$ s.t. $Q \Rightarrow \xrightarrow{x\vec{e}} \Rightarrow Q'$ and $P'\mathcal{S}Q'$;
- $P \xrightarrow{\overline{x}} (v\vec{y})\langle\vec{e}\rangle.P'$ implies $\exists Q'$ s.t. $Q \Rightarrow \xrightarrow{\overline{x}} (v\vec{y})\langle\vec{e}\rangle.Q''$, $Q'' \Rightarrow Q'$, and $P'\mathcal{S}Q'$;
- $P \xrightarrow{\langle\rho_1 \frown \cdots \frown \rho_n, \mathcal{R}\rangle} P'$ implies $\exists Q'$ s.t. $Q \Rightarrow \xrightarrow{\langle\rho_1, \mathcal{R}\rangle} \Rightarrow \cdots \Rightarrow \xrightarrow{\langle\rho_n, \mathcal{R}\rangle} \Rightarrow Q'$ and $P'\mathcal{S}Q'$.

If both $\mathcal{S}$ and its inverse are weak simulations then $\mathcal{S}$ is a weak bisimulation. Two processes $P$ and $Q$ are weakly equivalent, written $P \approx Q$, if $(P, Q)$ is in some weak bisimulation.

The last rule of Definition 6.4 extends classical weak bisimulation to continuous actions that are "chopped" by discrete $\tau$-actions. For example, $\texttt{wait}(3) \approx \texttt{wait}(1).\tau.\texttt{wait}(1).\tau.\tau.\texttt{wait}(1)$. The following theorem reveals that $\approx$ is an equivalence relation and itself a weak bisimulation.

THEOREM 6.5 (EQUIVALENCE AND WEAK BISIMULATION).
*(1) $\approx$ is an equivalence relation;*
*(2) $\approx$ is itself a weak bisimulation.*

PROOF. It is analogous to Theorem 6.2. □

## 6.3 Approximate Bisimulation

The notion of approximate bisimulation was first proposed in [20] to suit the study of hybrid systems. It allows the observations of two systems to be distant within a tolerable bound, rather than exactly identical. In [25], the approximate bisimulation was extended to allow precision not only between the observations of two systems, but also between the timely synchronisation between them. The work of [58] extends approximate bisimulation by allowing both a tolerance on time and value between the observations of two systems. It can be used to prove the approximate equivalence between a hybrid discrete-continuous model (in, e.g., Hybrid CSP) and the discrete SystemC [58] and C [53] code generated from it.

*Definition 6.6 (Approximate Bisimulation).* A binary relation $\mathcal{S}$ over the processes is an $(\epsilon, \delta)$-approximate simulation if for each $P\mathcal{S}Q$ we have:

- $P \xrightarrow{\tau} P'$ implies $\exists Q'$ s.t. $Q \Rightarrow Q'$ and $P'\mathcal{S}Q'$;
- $P \xrightarrow{x\vec{e}} P'$ implies $\exists Q'$ s.t. $Q \Rightarrow \xrightarrow{x\vec{e}} \Rightarrow Q'$ and $P'\mathcal{S}Q'$;
- $P \xrightarrow{\overline{x}} (v\vec{y})\langle\vec{e}\rangle.P'$ implies $\exists Q'$ s.t. $Q \Rightarrow \xrightarrow{\overline{x}} (v\vec{y})\langle\vec{e}\rangle.Q''$, $Q'' \Rightarrow Q'$, and $P'\mathcal{S}Q'$;
- $P \xrightarrow{\langle\rho, \mathcal{R}\rangle} P'$ implies $\exists Q'$ s.t. $Q \Rightarrow \xrightarrow{\langle\rho_1, \mathcal{R}_1\rangle} \Rightarrow \cdots \Rightarrow \xrightarrow{\langle\rho_n, \mathcal{R}_n\rangle} \Rightarrow Q'$, and $P'\mathcal{S}Q'$, where
$\|\text{Left}(\mathcal{G}) - \text{Left}(\mathcal{G}_1)\| \le \epsilon$ and $\big||\rho| - |\rho_1 \frown \cdots \frown \rho_n|\big| \le \delta$ with $\rho = \mathcal{A} \cdot \mathcal{G}$ and $\rho_1 = \mathcal{A}_1 \cdot \mathcal{G}_1$.

If both $\mathcal{S}$ and its converse are $(\epsilon, \delta)$-approximate simulations then $\mathcal{S}$ is an $(\epsilon, \delta)$-approximate bisimulation. $P$ and $Q$ are $(\epsilon, \delta)$-approximately bisimilar, written $P \approx_{\epsilon, \delta} Q$, if $(P, Q)$ is in some $(\epsilon, \delta)$-approximate bisimulation. In particular, the last rule says that, if two processes $P$ and $Q$ are in $(\epsilon, \delta)$-approximate bisimulation, then (1) the distance between observations of $P$ and $Q$ should not exceed $\epsilon$. The left states of guarantee-flows $\mathcal{G}$ and $\mathcal{G}_1$ of continuous variables defined by $P$ and $Q$ denote their observed states, respectively). Then, (2) the distance between two continuous actions of $P$ and $Q$ should not exceed $\delta$ (it compares the lengths of continuous actions). Finally, (3) the target processes $P'$ and $Q'$ should be in the same bisimulation.

Notice that weak bisimulation in Definition 6.4 is a $(0, 0)$-approximate bisimulation The distance allowed between states in approximate bisimulation can be considered as a kind of metrics in quantitative bisimulation [19, 28], which is however limited to states and traces. In both frameworks, its goal is to define a safe limit for the update period of the controller. The discretisation of hybrid model plays an important role in the model-based design of cyber-physical systems. To design a system combined with discrete, continuous and mobility, code at the implementation-level must determine how to sample data from the continuous plant, compute control commands, and perform the interactions between these different parts correctly. An efficient solution is to first have a hybrid control model that can be formally verified with safety guarantee, and then to discretise the continuous plant to obtain an algorithmic model to implement the cyber-physical system.

The notion of approximate bisimulation provides a bridge between hybrid models in control theory and discrete code in computer science. When a weak bisimulation relation between the hybrid model and its discretisation holds, the correctness of the code implementation is guaranteed. Actually, based on the approximate bisimulation of Definition 6.6, the discretisation of the whole **HpC** to correct executable code can be done similarly as for HCSP in [53, 58]. For example, consider the discretisation of the continuous evolution of time length $d > 0$. Concretely, the continuous process $\{\vec{e}_0 \mid \dot{\vec{v}} = \vec{e}\}_d \triangleq (\nu c)\{0, \vec{e}_0 \mid \dot{c} = 1, \dot{\vec{v}} = \vec{e} \,\&\, c < d\}$ evolves from $\vec{e}_0$ according to $\dot{\vec{v}} = \vec{e}$ for d time units. Its discretisation is given by the following theorem.

THEOREM 6.7 (DISCRETISATION). *Given a set of ODEs $\dot{\vec{v}} = \vec{e}$, with $\mathcal{N}(\vec{e}) \subseteq \mathcal{N}(\vec{v})$ and $\vec{e}_0$ as the initial value, satisfying the local Lipschitz condition on the interval $[0, d]$ for some $d > 0$, i.e., there exists a constant $L > 0$ such that $\|[\![\vec{e}\{\vec{c}_1/\vec{v}\}]\!] - [\![\vec{e}\{\vec{c}_2/\vec{v}\}]\!]\| \leq L\|\vec{c}_1 - \vec{c}_2\|$ for all $\vec{c}_1, \vec{c}_2 \in \mathbb{D}$, where $\mathbb{D} \subset \mathbb{R}^{|\vec{v}|}$ is any compact set. Let $\vec{c}$ be the values satisfying $\|\vec{c} - [\![\vec{e}_0]\!]\| \leq \epsilon'$, then for any precision $\epsilon > \epsilon' > 0$, there exists a step size $\delta > 0$, using coefficients $\Phi$ of the 4-stage Runge-Kutta method for computing approximated values of $\vec{v}$, such that $\mathrm{step}(\vec{y}, d) \triangleq (\nu c)\{0, \vec{y} \mid \dot{c} = 1, \dot{\vec{v}} = \vec{0} \,\&\, c < d\}$ and*

$$\{\vec{e}_0 \mid \dot{\vec{v}} = \vec{e}\}_d \quad \approx_{(\epsilon, \delta)} \quad \mu x(\vec{y}, z). \begin{pmatrix} [z \geq \delta].\mathrm{step}(\vec{y}, \delta).\overline{x}\langle \vec{y} + \Phi(\vec{y}, \delta) \cdot \delta, z - \delta\rangle \\ +[0 < z < \delta].\mathrm{step}(\vec{y}, z) + [z = 0].\mathtt{0} \end{pmatrix} @\langle \vec{c}, d\rangle$$

## 7 CASE STUDY: A NETWORK'S HANDOVER PROTOCOL

To put all the features and definitions of the **HpC** in motion, our case study is that of a handover protocol for next-generation's high-speed train control system. For the sake of clarity, we abstract away most of the design details and actual system parameters to only highlight key behaviours of the protocol, assuming wired or wireless private communications (i.e. not open broadcast radio channels, for obvious safety and security concerns).

### 7.1 Overview

The overall framework of the train control system is shown in Fig. 1. The network is composed of a train, of several railway-sectors, and of a terminus where the train stops. The train starts from the departure and ends at the terminus, under the coordinated control of the consecutive sectors.

The railway sector can directly control the train by sensing and actuating using private channels of the train (its ready-set interface). These control links (channels) are transferred as messages between adjacent sectors to handover the train from one to the other. The sector receiving the channels takes over control of the train until the train leaves it, in which case a handover occurs. The network may of course be extended to contain more sectors, bifurcations, terminals and trains but, for simplicity, we just require one sector to control at most one train at a time. The status of a sector is hence either *free* or *busy*. All sectors are initially free. As Fig. 1 shows, if a sector is free (a), i.e., it handles no train, then it will keep on stand-by, monitoring requests from its neighbour sector(s). Once it receives a handover request, it switches to the busy status and gets ready to take control of the approaching train. If the sector is busy, on the contrary, it will reply to its neighbour to refuse the handover request. It will return to the free status once the train it controls leaves the current sector. As illustrated by (c) and (d) in Fig. 1, each sector has a handover point, near the end point of the sector. Once the train reaches the handover point of a sector, the sector will raise a handover request to the next sector. If the handover request is accepted (c), then it will transfer the channels connecting to the train to the next sector. Otherwise (d), it will decelerate the train to prevent it from entering the adjacent sector, and then try to hand it over again. The terminus can be regarded as a special sector that decelerates the train until it becomes immobile i.e., "terminates".

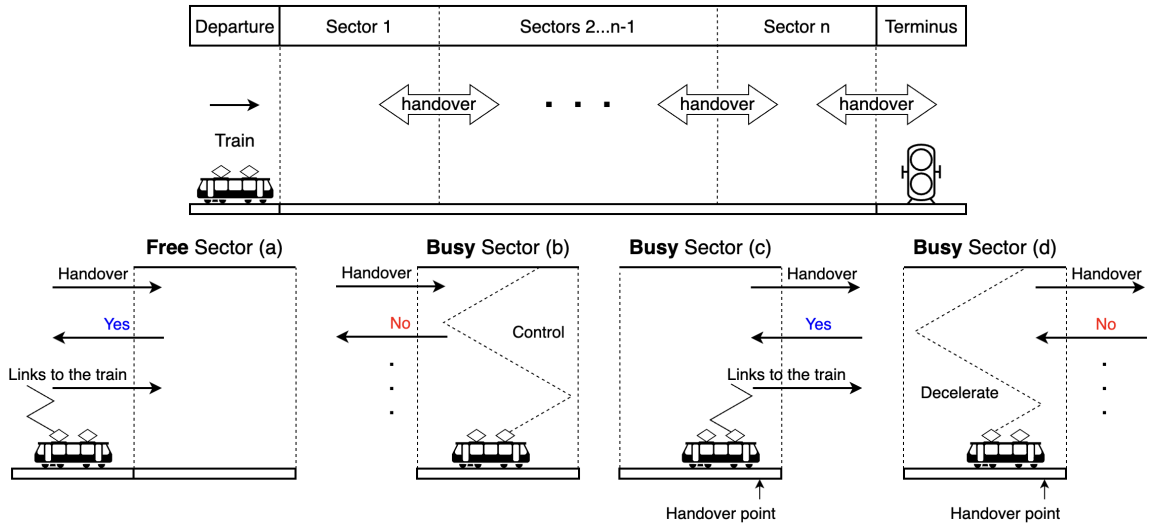

Fig. 1. Scenarios of the handover protocol

### 7.2 Trains

The behaviour of a collection of (potentially infinite) trains is modelled by the following replication.

$$\text{TRAIN} \triangleq \ !train(handover', yes', no', switch).\overline{handover'}.$$
$$\begin{pmatrix} yes'.(vp, v, a, q)\overline{switch}\langle p, v, a, q\rangle. \\ \left\{ \begin{array}{ccc} 0 & | & \dot{p} = v \\ 0 & | & \dot{v} = a \\ 0 & | & \dot{a} = 0 \\ 0 & | & \dot{q} = 0 \end{array} \& p < q, \{\overline{p}, \overline{v}, a, q\} \right\} \\ + \\ no'.\overline{\texttt{wait}(10)}.\overline{train}\langle handover', yes', no', switch\rangle \end{pmatrix}$$

Once TRAIN is invoked by channel *train*, it forks an instance of a train process which starts a handover request by channel *handover'* to the (first) sector to control the train. If the request is replied by *yes'*, the channels of the train are created, i.e., $(vp, v, a, q)$. The newly created private channels will be sent to the controlling sector by channel *switch*, and then the train will start running according to the ODE under the control of the sector. However, if the request is refused (*no'*), the train will wait for 10s before retrying. The position, velocity, acceleration, and the destination of a train are denoted by names $p$, $v$, $a$, and $q$ with 0s as initial values. A railway sector controls a train by sensing its real-time position ($p$) and velocity ($v$) and actuating the acceleration ($a$) and destination ($q$), according to the interface $\{\overline{p}, \overline{v}, a, q\}$. The boundary condition $p < q$ means that a train terminates once it reaches the destination. Note that the value of the destination can be modified by "actuation" ($q \in \{\overline{p}, \overline{v}, a, q\}$).

### 7.3 Sectors

The sectors control the evolution of the train by sensing the real-time position and velocity and by actuating the acceleration of the train. A sector is either *free* or *busy*. A sector is free if it controls no train and busy otherwise. Thus, a free sector accepts the handover request from its prior sector while a busy sector refuses.

*7.3.1 Behaviours.* Let $\vec{\xi}$ = *handover, yes, no, reset*, the behaviour of a free sector takes two channels *free* and *busy* as parameters to switch the sector from state free to state busy.

$$\text{Free} \triangleq free(\vec{\xi}).handover.(\overline{yes} \| \overline{busy}\langle\vec{\xi}\rangle)$$

A Free sector is invoked by *free* and receives the channels (*handover, yes,* and *no*) from its predecessor and the channel to *reset* its state back to *free* (as explained later). Once invoked, it monitors *handover* requests from its predecessor. Once it receives such a request, it (1) notifies its predecessor sector that the request is accepted by *yes*, and (2) get into the busy state by invoking *busy*. These two actions execute in parallel. A collection of FREE sectors can be modelled by replication: FREE $\triangleq$ !Free. A free sector becomes busy once it receives the handover request from its predecessor, and the behaviour of a busy sector is modelled by

$$\text{Busy} \triangleq busy(\vec{\xi}).\left(handover.(\overline{no} \| \overline{busy}\langle\vec{\xi}\rangle) + reset.\overline{free}\langle\vec{\xi}\rangle\right)$$

The mirror process Free takes channels *free* and *busy* as parameters. A Busy sector is invoked by channel *busy* and receives the channels (*handover, yes,* and *no*) from its predecessor and channel *reset* to flip its state to free. After the invocation, it either receives a *handover* request from its predecessor or it is *reset*. In the first case, it refuses the handover request by *no* and, meanwhile, stay in the busy state by invoking *busy* again. In the other case, it returns to state free by invoking channel *free*. Notice that the choice between *handover* and *reset* is non-deterministic: if the choice cannot be made at the moment of the interruption, the sum it forms is delayed (by application of rule [Wait]). A collection of BUSY sectors is also defined by replication BUSY(*busy, free*) $\triangleq$ !Busy(*busy, free*). Once a free sector accepts the handover request from its neighbour, it switches to the busy state (see Free) and starts to control the approaching train. The controller's behaviour is modelled by

$$\text{Control} \triangleq ctrl(\vec{\eta}).channels(p, v, a, q).\overline{q}\langle p_e\rangle.\overline{comp}\langle\vec{\eta}, p, v, a, q\rangle$$

where $\vec{\eta} = (p_s, p_h, p_e, d, handover', yes', no', switch, channels, reset)$ are defined as follows. Parameters $p_s$, $p_h$, and $p_e$ denote the start, handover, and end points of the sector. The sector sends handover requests to the next sectors via channel *handover'*, and receives successful and failed responses from channels *yes'* and *no'*, respectively. Duration $d$ is the control period. The sector resets its own state by channel *reset*. If the handover request to the next sector is accepted, the current sector sends the private channels ($p$, $v$, $a$, and $q$) of the leaving train to the next sector by *switch*. Symmetrically, if the sector accepts the handover request from the previous sector, it receives the channels of the approaching train on *channels*. Process Control is invoked by $ctrl(\vec{\eta})$. Then, it waits for the parameters ($p$, $v$, $a$, and $q$) of the train approaching from the previous sector, modifies the destination of the train to the end point of the current section, i.e., $\overline{q}\langle p_e\rangle$, and finally invokes the computing function *comp* below. We can use replication to treat Control as a function called repeatedly by process CTRL $\triangleq$ !Control. The computing behaviour of a (busy) sector is invoked by channel *comp*. Then, it senses the real-time position $p_0$ of the train it controls by the position name $p$ of the train, i.e., $p(p_0)$. If the train has not reached the handover point of the sector (i.e. $p_0 < p_h$), then the sector continues to sense the real-time velocity of the train, based on which a control command is computed by the function $f$, specified in Section 7.6.5. After actuating the acceleration of the train by $\overline{a}\langle f(p_0, v_0, p_e, d)\rangle$, the sector waits for one period, i.e., wait($d$), before invoking the next computation. If the train crosses the handover point ($p_0 \geq p_h$), then the sector sends a handover request to the next sector by $\overline{handover'}$ and wait for a reply. If the reply is $no'$, it continues as usual. When it receives $yes'$, it sends the private channels of the train to the next sector, i.e., $\overline{switch}\langle p, v, a, q\rangle$, and then monitors the real-time position of the train until it leaves the sector ($p < p_e$ becomes **false**). After that, it *reset*s its state to free and restart Control process by *ctrl*. Process Compute can also be treated as a function COMP ≜ !Compute.

$$\begin{aligned}
\text{Compute} \triangleq\ & comp(\vec{\eta}, p, v, a, q).p(p_0). \\
& \begin{pmatrix}
\textbf{if } p_0 < p_h \textbf{ then} \\
\quad v(v_0).\overline{a}\langle f(p_0, v_0, p_e, d)\rangle.\text{wait}(d).\overline{comp}\langle \vec{\eta}, p, v, a, q\rangle \\
\textbf{else} \\
\quad \overline{handover'}. \\
\quad \begin{pmatrix} yes'.\overline{switch}\langle p, v, a, q\rangle.(vc)\{0 \mid \dot{c} = 1\ \&\ p < p_e\}.\overline{reset}.\overline{ctrl}\langle \vec{\eta}\rangle \\ + \\ no'.v(v_0).\overline{a}\langle f(p_0, v_0, p_e, d)\rangle.\text{wait}(d).\overline{comp}\langle \vec{\eta}, p, v, a, q\rangle \end{pmatrix}
\end{pmatrix}
\end{aligned}$$

*7.3.2 Sector Creation.* In Section 7.3.1, the behaviours of sectors consists of four replicated elements in two parts: FREE and BUSY manage the states of sectors, and CTRL and COMP specify the control procedure. The following process initializes a (free) sector:

$$\text{Sector} \triangleq sec(\vec{\xi}', \vec{\eta}').(\nu\ reset)\left(\overline{free}\langle\vec{\xi}\rangle \| \overline{ctrl}\langle\vec{\eta}\rangle\right)$$

We write $\vec{\xi}' = \vec{\xi} \setminus reset$ and $\vec{\eta}' = \vec{\eta} \setminus reset$ for removing the name *reset* from $\vec{\xi}$ and $\vec{\eta}$. The sequences of names $\vec{\xi}$ and $\vec{\eta}$ have been defined in Section 7.3.1. Process Sector is invoked by channel *sec* with receiving $\vec{\xi}', \vec{\eta}'$. Then, it calls FREE via channel *free* and forks a process to manage the state of the sector (initially *free*). Meanwhile, it calls CTRL via channel *ctrl* to start the control procedure. These two processes operate in parallel and are connected via the local channel *reset*. If the train leaves the current sector, the control procedure will reset the state of the sector to free (see Busy and Compute). We consider potentially infinite sectors, thus we define the replication of sectorsSECT ≜ !Sector. Sectors can be defined by the parallel composition to create a new (free) sector by channel *sec*:

$$\begin{aligned}
\text{SECTOR} \triangleq\ & \text{SECT} & \text{(Call Entry)} \\
& \|\ \text{FREE} \| \text{BUSY} & \text{(State)} \\
& \|\ \text{CTRL} \| \text{COMP} & \text{(Function)}
\end{aligned}$$

For example, given the sequences of names $\vec{\xi}'$ and $\vec{\eta}'$ above, $\vec{\eta}'' = \vec{\eta}'\{10/p_s, 14/p_h, 15/p_e, 1/d\}$ and $\overline{sec}\langle\vec{\xi}', \vec{\eta}''\rangle\|\text{SECTOR}$ create a free sector whose start ($p_s$), handover ($p_h$), and end ($p_e$) points are at 10 km, 14 km, and 15 km, respectively, and the control period ($d$) of the sector is 1 s.

## 7.4 Terminus

The terminus can be regarded as a particular sector which always stops trains. The behaviour of terminus can be defined using replication: Once it is invoked by *term*, it will fork a recursion that refuses ($\overline{no}$) any *handover* request it receives.

$$\text{TERMINUS} \triangleq\ !term(handover, no).\mu x.handover.\overline{no}.x$$

### 7.5 Network

The resource of the whole system is composed of trains, sectors, and a terminus:

$$\text{RESOURCE} \triangleq \text{TRAIN} \| \text{SECTOR} \| \text{TERMINUS}$$

We create channels $handover_i$, $yes_i$, $no_i$, and $channels_i$ to connect the $i$-th and the $(i+1)$st sectors. The $i$-th sector sends handover requests to the next sector via channel $handover_i$ and it will be replied by $yes_i$ or $no_i$. If $yes_i$, it then sends the channels of the train to the next sector via channel $channels_i$. The network contains three consecutive sectors, a terminus (which could be treated as the fourth sector) and a train is placed at the start point of the first sector, initially.

Let $\vec{x}_i = handover_i, yes_i, no_i$ for $i \in \{0, 1, 2, 3\}$ and $\vec{y} = channels_0, channels_1, channels_2, channels_3$. The complete train control Network is modelled as follows:

$$\text{Network} \triangleq (\nu\, train, sec, free, busy, ctrl, comp, term, stop, brake)(\text{Starter} \| \text{RESOURCE})$$

$$\text{Starter} \triangleq (\nu\vec{x}_0, \vec{x}_1, \vec{x}_2, \vec{x}_3, \vec{y}) \begin{pmatrix} \overline{train}\langle \vec{x}_0, channels_0 \rangle \| \\ \overline{sec}\langle \vec{x}_0, 0, 4, 5, 1, \vec{x}_1, channels_0, channels_1 \rangle \| \\ \overline{sec}\langle \vec{x}_1, 5, 9, 10, 1, \vec{x}_2, channels_1, channels_2 \rangle \| \\ \overline{sec}\langle \vec{x}_2, 10, 14, 15, 1, \vec{x}_3, channels_2, channels_3 \rangle \| \\ \overline{term}\langle handover_3, no_3 \rangle \end{pmatrix}$$

The Starter creates all the physical entities (a train, three sectors, and a terminus) and connects them together through links (channels) $\vec{x}_i$. In addition, from the actual parameters (configurations) of these entities, we can see that the period of each sector and the terminus is 1 s, the length of each sector is 5 km with the handover point 1 km to the end point.

### 7.6 Verification of the Handover Protocol Scenarios

This section focuses on the verification of the most critical component of the railway network's control: the handover protocol. The specification of the complete model of the network is included in the supplementary materials which will be shared online.

Unlike the complete case study, which features dynamicity, mobility, orchestration, the network considered in the present scenario contains only one train Train and two sectors, Left and Right. Its purpose is to verify the robustness of the handover protocol with respect to physical models and observations manifesting bounded physical disturbances and, nonetheless, prove the approximate bisimulation between an abstract, perfect model, and the realistic model with noise/disturbance.

The handover protocol is verified against its two possible scenarios: a successful handover, next, and a failed one. The train is placed at the starting point of the left sector initially. Under the control of the two sectors, the train will run to the right and finally terminate at the end of the right sector. By using replication, it can be extended to a network containing potentially infinite trains, controllers, terminus and sectors, which can be created and terminated dynamically.

*7.6.1 The Train Model.* Let $u$ represent an arbitrary disturbance of the environment on the train and let constant len = 5 km denote the length of each sector. In the real world, however, the run of a train will be disturbed by many factors such as wind, drift, heat, friction and slope. All these disturbances are abstracted as an external and uncontrolled input $u$ in the model. We assume that $u$ is free to change dynamically within certain bounds $[-0.1, 0.1]$. The Train first creates three private variables $(\nu p, v, a)$ to denote its position $p$, velocity $v$, and acceleration $a$ of the train. These variables are sent out via $channels$ by $\overline{channels}\langle p, v, a \rangle$. The sector receiving them takes control of the train. The train runs according to the set of ODEs specified in Run. The boundary condition $p < 2\text{len}$ means that the length of the whole journey of the train will be $2\text{len} = 10$ km long. The ready set (interface) $\{\overline{p}, \overline{v}, a\}$ indicates that the position ($p$) and velocity ($v$) of the train are sensors while the acceleration ($a$) can be actuated by the sector controlling the train. Once the train reaches the end of the right sector, it sends its last position (2len) out before termination.

$$\begin{aligned} \text{Train} &\triangleq (\nu p, v, a)(\overline{channels}\langle p, v, a\rangle.\text{Run} \parallel \text{Observer}) \\ \text{Run} &\triangleq \left\{ \begin{array}{l} 0 \mid \dot{p} = v \\ 0 \mid \dot{v} = a + u \\ 0 \mid \dot{a} = 0 \end{array} \& p < 2\text{len}, \{\overline{p}, \overline{v}, a\} \right\}.\overline{p}\langle 2\text{len}\rangle \\ \text{Observer} &\triangleq \{0 \mid \dot{x} = v \& p < 2\text{len}\}(x').\{x' \mid \dot{x} = 0\} \end{aligned}$$

Process Observer monitors the evolution of the train. It calculates its location using a free variable $x$ by sensing the recorded velocity $v$ of the train defined in process Run and without actuating the dynamic of the train. Once the evolution of the train terminates, the physical variables $p$, $v$, $a$ are no longer visible. We hence sustain $x$ to its latest value once the train reaches the end of the right sector ($p = 2\text{len}$), as specified by the ODE $\dot{x} = 0$.

The dynamic of the Train is not affected . To prove bisimulation properties, however, it is necessary to use the Observer to track the position of the train, because the evolution of $p$ is private to the Train (variable $p$ is under the scope of a restriction $\nu$). The variable $x$ in the Observer is used to mimic or simulate the evolution of $p$ in Run.

*7.6.2 The Left Sector.* The Left sector first takes control of the train by receiving the physical variables of the train, i.e., $channels(p, v, a)$, and then starts its control Loop. Let us highlight the constants for the handover point $\text{p}_h = 4$ km, the track length len = 5 km, and the update period of the sector's controller d = 1 s. At each iteration of the Loop, the sector first reads the current position of the train, i.e., $p(p_0)$. If the train has not reached the handover point ($p_0 < \text{p}_h$), the sector will continue to read the current velocity, i.e., $v(v_0)$, compute a safe acceleration $f(p_0, v_0, \text{len}, \text{d})$ based on the sensed data, and then actuate the train with the new acceleration. The function $f : \mathbb{R}^4 \to \mathbb{R}$ reflects the control strategy and will be specified in Section 7.6.5. After that, the sector waits for d time units before starting the next iteration of control, i.e., $\text{wait}(\text{d}).\overline{ctrl}$.

However, if the train has crossed the handover point ($p_0 \geq \text{p}_h$), the sector will send a handover request to the right sector ($\overline{handover}$) and wait for a reply. If the reply is *no*, i.e., the handover is refused, it will continue controlling the train to decelerate it. If the reply is *yes*, the left sector sends the physical variables of the train to the right sector ($\overline{switch}\langle p, v, a\rangle$), switching the control of the train from the left to the right sectors.

$$\begin{aligned} \text{Left} &\triangleq channels(p, v, a).\text{Loop} \\ \text{Loop} &\triangleq \mu\, ctrl.p(p_0).\begin{pmatrix} \textbf{if } p_0 < \text{p}_h \textbf{ then} \\ \quad v(v_0).\overline{a}\langle f(p_0, v_0, \text{len}, \text{d})\rangle.\text{wait}(\text{d}).\overline{ctrl} \\ \textbf{else} \\ \quad \overline{handover}.\begin{pmatrix} yes.\overline{switch}\langle p, v, a\rangle \\ + \\ no.v(v_0).\overline{a}\langle f(p_0, v_0, \text{len}, \text{d})\rangle.\text{wait}(\text{d}).\overline{ctrl} \end{pmatrix} \end{pmatrix} \end{aligned}$$

*7.6.3 The Right Sector: Successful Handover.* In this scenario, since the Right sector contains no train initially, it should first reply $\overline{yes}$ to the *handover* request of the left sector. Then, it takes the control of the coming train by $switch(p, v, a)$ and starts the control Loop'. At each iteration of Loop', the right sector checks whether the train has reached the endpoint. If not ($p_0 < 2\text{len}$), it drives the train periodically according to the algorithm specified by $f$; otherwise, control terminates.

$$\begin{aligned} \text{Right} &\triangleq handover.\overline{yes}.switch(p, v, a).\text{Loop}' \\ \text{Loop}' &\triangleq \mu\, ctrl.p(p_0).\textbf{if } p_0 < 2\text{len} \textbf{ then } v(v_0).\overline{a}\langle f(p_0, v_0, 2\text{len}, \text{d})\rangle.\text{wait}(\text{d}).\overline{ctrl} \textbf{ else } 0 \end{aligned}$$

REMARK 4. *The physical variables $p$, $v$, and $a$ in* Run *depict the continuous evolution of the position, velocity, and acceleration of the* Train. *Meanwhile, they can be used as channels by sectors to control the train. In this sense, they represent the "control" of the train: any sector obtaining these variables takes control of the train. In particular, these variables can be exchanged between sectors, indicating the handover of control, and reflecting the "mobility" of the system. In addition, this example reflects the convention of the classic $\pi$-calculus that "names are the first-class citizen".*

*7.6.4 The Right Sector: Failed Handover.* In this scenario, the right sector always refuses the handover request from the left. The behaviour of the right sector can be modelled by

$$\text{Right}' \triangleq \mu x.handover.\overline{no}.x$$

It replies $\overline{no}$ whenever it receives a *handover* request.

*7.6.5 Control Algorithm.* We mentioned in Section 7.6.2 that a sector controls a train by using a control command function $f : \mathbb{R}^4 \to \mathbb{R}$. Given a real-time position $p_0$, velocity $v_0$, end point $p_e$ and period $d$ of the sector, function $f$ computes the acceleration $a$ of the train. Let $v_{\max} = 40$ m/s be the maximum velocity of the train and $a_{\min} = -1$ m/s$^2$ the braking deceleration of the train. The control performed by $f$ uses the Maximum Protection Curve [57] and is computed as follows:

$$V_{\lim}(p_0, p_e) = \begin{cases} v_{\max} & \text{if } p_e - p_0 \geq \frac{v_{\max}^2}{(-2a_{\min})} \\ \sqrt{-2a_{\min} \cdot (p_e - p_0)} & \text{if } 0 < p_e - p_0 < \frac{v_{\max}^2}{(-2a_{\min})} \end{cases}$$

If the end point is farther than the minimal safe distance $-v_{\max}^2/(2a_{\min})$ of the train, the upper limit velocity of the vehicle can be the maximum $v_{\max}$. If not, the velocity should not exceed $\sqrt{-2a_{\min} \cdot (p_e - p_0)}$ in order not to enter the next sector. At each iteration, the control algorithm predicts the position $p'$ and velocity $v'$ of the train at the next period based on the maximum acceleration $a_{\max} = 1$ m/s$^2$. Concretely, they can be computed by $v' = v_0 + a_{\max} \cdot d$ and $p' = p_0 + v_0 \cdot d + \frac{1}{2} \cdot a_{\max} \cdot d^2$, where $d > 0$ is the update period of the sector. If, at the next period, the velocity does not exceed the upper limit computed as above, i.e., $v' \leq V_{\lim}(p', p_e)$, then the maximum acceleration $a_{\max}$ is safe. If not, it continues to test if a constant velocity (no acceleration or deceleration) is safe, i.e., $v_0 \leq V_{\lim}(p_0 + v_0 \cdot d, p_e)$. If not, the train brake with the minimal deceleration $a_{\min}$. The above control strategy can be summarised as:

$$f(p_0, v_0, p_e, d) = \begin{cases} a_{\max} & \text{if } v' \leq V_{\lim}(p', p_e) \\ 0 & \text{if } v_0 \leq V_{\lim}(p_0 + v_0 \cdot d, p_e) \\ a_{\min} & \text{otherwise} \end{cases}$$

*7.6.6 Systems.* The System of the successful scenario (Section 7.6.3) is modelled by the composition of the Train and Left and Right sectors, and they communicate with each other through local channels (of scope restricted by $\nu$):

$$\text{System} \triangleq (\nu \, link, handover, switch, yes, no)(\text{Train}\|\text{Left}\|\text{Right})$$

Channel *link* connects the Train and the Left sector. The other channels (*handover, switch, yes,* and *no*) connect the Left and Right sectors. Similarly, the System' of the failed scenario (Section 7.6.4) is as follows:

$$\text{System}' \triangleq (\nu \, link, handover, switch, yes, no)(\text{Train}\|\text{Left}\|\text{Right}')$$

*7.6.7 Approximate Bisimulations.* In the successful scenario, the train runs into the right sector smoothly and finally stops at the end of the right sector. If there is no disturbance, the expected behaviour of the train can be described by the following specification (SPEC):

$$\begin{aligned}
\text{SPEC} &\triangleq (\nu p, v)\big((\text{Up}.\text{Stable}.\text{Down}) \| \text{Observer}\big) \\
\text{Up} &\triangleq \{0, 0 \mid \dot{p} = v, \dot{v} = 1 \,\&\, p < 0.8 \text{ km}\}(p_1, v_1) \\
\text{Stable} &\triangleq \{p_1, v_1 \mid \dot{p} = v, \dot{v} = 0 \,\&\, p < 9.2 \text{ km}\}(p_2, v_2) \\
\text{Down} &\triangleq \{p_2, v_2 \mid \dot{p} = v, \dot{v} = -1 \,\&\, p < 10 \text{ km}\}
\end{aligned}$$

Process Observer is defined in Section 7.6.1. According to this specification, the evolution of the (undisturbed) train is composed of the following three stages:

**Up** The train speeds up from the initial state ($p = 0$ m and $v = 0$ m/s) with the maximum acceleration 1 m/s$^2$ for 0.8 km when it reaches the maximum velocity 40 m/s.

**Stable** Then, the train runs with constant velocity (40 m/s) until it is 0.8 km close to the end of the right sector ($10 - 0.8 = 9.2$ km).

**Down** Finally, the train decelerates with the minimum acceleration $-1$ m/s$^2$ until it stops at the end of the right sector (10 km).

The specification (SPEC) describes the ideal model of the train control system without disturbance. The following proposition reveals that the disturbed System model (Section 7.6.6) is $(0.4 \text{ km}, 0)$-approximately bisimilar to the SPEC model, indicating that the distance between the disturbed and undisturbed trains is within 0.4 km. By simulation, the maximum distance is around 0.205 km.

Proposition 7.1. SPEC $\approx_{(0.4 \text{ km}, 0)}$ System.

Proof. We first prove that the distance between the real-time positions of the trains in SPEC (undisturbed) and System (disturbed) is not larger than 0.4 km, based on which we then prove that the following relation forms a $(0.4 \text{ km}, 0)$-approximate bisimulation. More intuitively, it can be represented by the hybrid automaton in the left of Fig. 3, where each state only observes variable $x$.

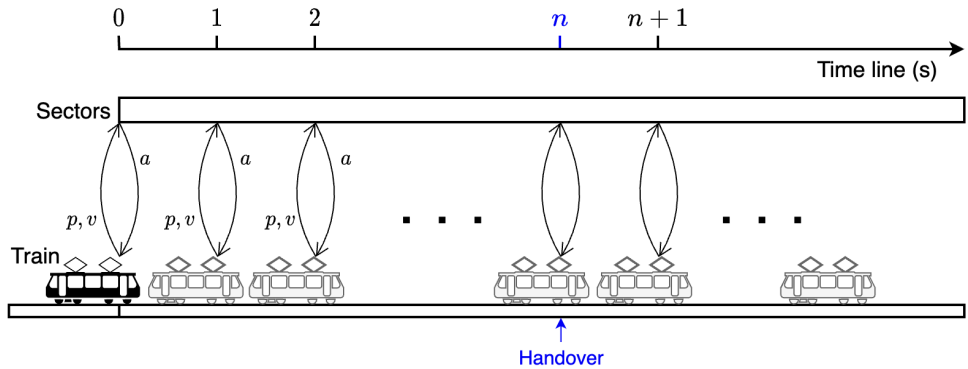

Fig. 2. The execution of System

The execution of the System model can be depicted by Fig. 2. The sector senses the position ($p$) and velocity ($v$), computes a new acceleration ($a$) based on the sensed data according to the algorithm specified in Section 7.6.5, and actuates the train with the computed acceleration every 1 second (the control period of the sector). Here the communication and computation are represented by a sequence of $\tau$-actions, which are timeless. Once the Left sector senses that the train has reached the handover point, say at time $n$, it raises a handover request, which will be accepted by the Right sector, and then the right sector takes over the control. Since the handover is successful and both sectors adopt the same control algorithm (Section 7.6.5), the handover of the train is seamless and the train runs into the right sector without realising any difference.

$$\mathcal{S} = \{(\text{SPEC}, \text{System})\} \cup \left\{ (P, Q) \middle| \begin{array}{l} \text{SPEC} \xrightarrow{\langle \rho_1, \emptyset \rangle} \cdots \xrightarrow{\langle \rho_n, \emptyset \rangle} P \\ \text{System} \Rightarrow \xrightarrow{\langle \rho'_1, \emptyset \rangle} \Rightarrow \cdots \Rightarrow \xrightarrow{\langle \rho'_m, \emptyset \rangle} \Rightarrow Q \end{array}, |\rho_1 \frown \cdots \frown \rho_n| = |\rho'_1 \frown \cdots \frown \rho'_m| \right\}$$

The handover procedure can also be described by a sequence of $\tau$-actions and hence it is timeless. After that, under the control of the right sector, the train runs and finally terminates at the end point of the sector. Intuitively, the execution of System can be depicted by the hybrid automaton on the right of Fig. 3, where

$$\begin{array}{rl} B_1 \triangleq & v + a_{\max} \cdot d \leq V_{\lim}(p + v \cdot d + \frac{1}{2} \cdot a_{\max} \cdot d^2, 2\text{len}) \\ B_2 \triangleq & v \leq V_{\lim}(p + v \cdot d, 2\text{len}) \end{array}$$

indicate that the acceleration 1 m/s² and the constant velocity are safe at the next period, respectively. We only observe the variable $x$ of the state here (in red).

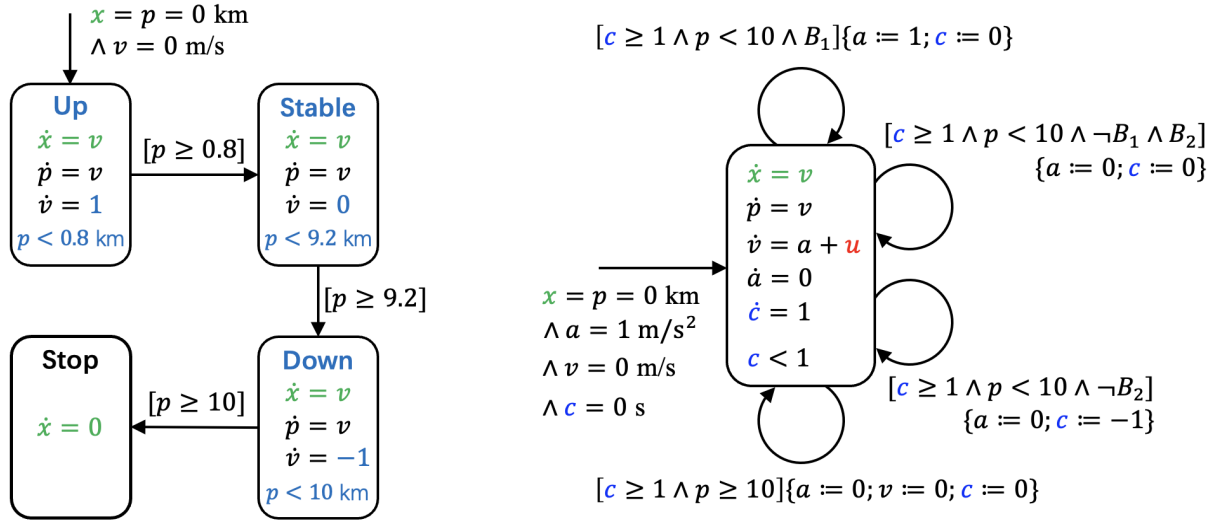

Fig. 3. The hybrid automata of SPEC (left) and System (right)

In order to observe the difference between the real-time positions ($x$) of the trains in SPEC and System, we compose the hybrid automata in Fig. 3, resulting in the composition shown in Fig. 4, where we subscript the variables of SPEC by 1 and System by 2 to distinguish between them.

Specifically, the composed hybrid automaton $\mathcal{H}$ consists of two independent components: one representing the evolution of SPEC variables ($x_1, p_1, v_1, a_1, c_1$) and the other representing System variables ($x_2, p_2, v_2, a_2, c_2$). Here, we introduce two fresh variables, $a_1$ and $a_2$, representing acceleration, to unify the dynamics of both components. For example, the transition from **Up** to **Stable** in the SPEC is described by the reset operation $a_1 := 0$ in $\mathcal{H}$, when the guard condition $[p_1 \geq 0.8 \wedge c_1 \geq 1]$ is met. The clock $c_1$, which is implicit in the SPEC, is introduced to force the transitions to take place at each time step. Consequently, proving the desired bisimulation relation is equivalent to showing that trajectories of the automaton $\mathcal{H}$ will never reach the unsafe region $|x_1 - x_2| \geq 0.4$ km. The latter problem is a typical *safety verification problem*: proving that a hybrid automaton, starting from an initial set of states, will never enter an unsafe region over an infinite-time horizon.

A well-established method to solve safety verification problem of hybrid systems is that of using *differential invariants*. Given a hybrid automaton, a differential invariant is a subset of the state space that remains invariant under the system's continuous dynamics, i.e., all trajectories starting from the differential invariant remain within it forever. To prove system safety, it suffices to find a differential invariant that includes the initial states and excludes the unsafe region. For the given problem, we can prove the existence of such a differential invariant that witnesses the safety of $\mathcal{H}$ with respect to the unsafe region $|x_1 - x_2| \geq 400$, implying that the relation $\mathcal{S}$ is a $(0.4 \text{ km}, 0)$-approximate bisimulation, i.e., SPEC $\approx_{(0.4 \text{ km}, 0)}$ System.

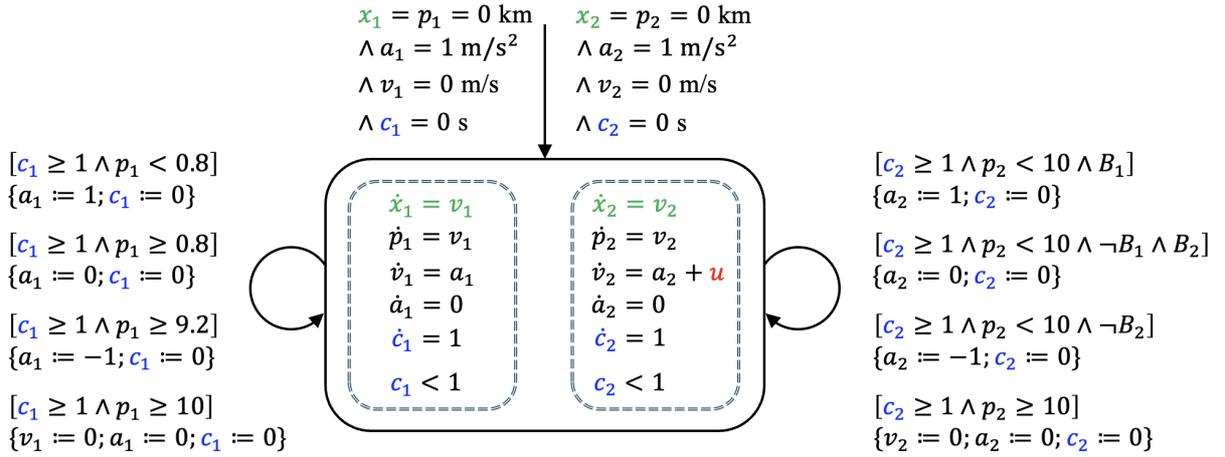

Fig. 4. The composed hybrid automaton $\mathcal{H}$

The existence of the aforementioned differential invariant is established by using *barrier certificates* [26, 43]. However, a detailed discussion of barrier certificates is beyond the scope of this paper. Appendix B provides formal definitions of hybrid automata and barrier certificates, along with an explanation of how to compute the desired differential invariant using these techniques. □

In the failed scenario, the expected behaviour of the undisturbed train can be described by the following specification (SPEC′), depicting a train stopping at the endpoint of the left sector (5 km):

$$\begin{aligned}
\text{SPEC}' &\triangleq (\nu p, v)\big((\text{Up}'.\text{Stable}'.\text{Down}')\|\text{Observer}\big) \\
\text{Up}' &\triangleq \{0, 0 \mid \dot{p} = v, \dot{v} = 1 \ \& \ p < 0.8 \text{ km}\}(p_1, v_1) \\
\text{Stable}' &\triangleq \{p_1, v_1 \mid \dot{p} = v, \dot{v} = 0 \ \& \ p < 4.2 \text{ km}\}(p_2, v_2) \\
\text{Down}' &\triangleq \{p_2, v_2 \mid \dot{p} = v, \dot{v} = -1 \ \& \ p < 5 \text{ km}\}
\end{aligned}$$

In the same (mirror) way as above, we can prove the following result:

PROPOSITION 7.2. SPEC′ $\approx_{(0.3 \text{ km}, 0)}$ System′.

## 8 CONCLUSIONS

In this article, we presented the definition of a compositional and expressive extension of the $\pi$-calculus with hybrid, discrete, real-timed and continuous processes, i.e., the hybrid $\pi$-calculus (**HpC**). The **HpC** allows to express and capture complex verification issues posed by dynamic, mobile and hybrid systems. We lifted the theory of bisimulation and congruence of the $\pi$-calculus to the **HpC**. In addition, we illustrate the expressive power, modelling and verification capabilities of the **HpC** by considering a functionally realistic model of a parametric hybrid (discrete/continuous) protocol: the handover protocol of a modular train control system. As a result, we obtain the first characterisation of process calculus that has the capability to compositionally orchestrate the interaction of discretely timed processes and continuously timed ordinary differential equations (ODEs) in the presence of dynamicity and mobility. All items of the **HpC** are first-class citizens:

- Discrete and continuous processes are equal hybrid processes.
- Discrete channels and continuous variables are equal hybrid names.
- All channels and processes, discrete or continuous, have meaning in the discrete and continuous operational semantics, providing the first hybrid semantics for the $\pi$ calculus.

The **HpC** supports control-theoretic approximate bisimulation for a process calculus with dynamic channels and processes.

**Future works**

The **HpC** and its case study build the foundation to study the verification of pervasive internet(s) of hybrid things. Our future works focus on developing three aspects to empower its practicality:

(1) Establish a proof system for the **HpC** based on the generalised hybrid Hoare logic [59];
(2) Extend the **HpC** to a hybrid session $\pi$-calculus with a powerful session type system [32, 46];
(3) Generate executable code automatically from **HpC** processes, the correctness of which (discretisation) will be proved using approximate bismulation (Definition 6.6).

## REFERENCES

[1] Rajeev Alur, Costas Courcoubetis, Thomas A. Henzinger, and Pei-Hsin Ho. 1992. Hybrid Automata: An Algorithmic Approach to the Specification and Verification of Hybrid Systems. In *Hybrid Systems (LNCS, Vol. 736)*. Springer, 209–229.
[2] Bogdan Aman and Gabriel Ciobanu. 2013. Real-Time Migration Properties of rTiMo Verified in Uppaal. In *SEFM 2013 (LNCS, Vol. 8137)*. Springer, 31–45.
[3] MOSEK ApS. 2019. *MOSEK Optimizer API for Julia. Version 10.1.13*. https://docs.mosek.com/latest/juliaapi/index.html
[4] Jos C. M. Baeten and Jan A. Bergstra. 1997. Process Algebra with Propositional Signals. *Theor. Comput. Sci.* 177, 2 (1997), 381–405.
[5] Jos C. M. Baeten and Cornelis A. Middelburg. 2002. *Process Algebra with Timing*. Springer.
[6] Richard Banach. 2024. Core Hybrid Event-B III: Fundamentals of a reasoning framework. *Sci. Comput. Program.* 231 (2024), 103002.
[7] Richard Banach, Michael J. Butler, Shengchao Qin, Nitika Verma, and Huibiao Zhu. 2015. Core Hybrid Event-B I: Single Hybrid Event-B machines. *Sci. Comput. Program.* 105 (2015), 92–123.
[8] Richard Banach, Michael J. Butler, Shengchao Qin, and Huibiao Zhu. 2017. Core Hybrid Event-B II: Multiple cooperating Hybrid Event-B machines. *Sci. Comput. Program.* 139 (2017), 1–35.
[9] Albert Benveniste, Benoît Caillaud, Dejan Nickovic, Roberto Passerone, Jean-Baptiste Raclet, Philipp Reinkemeier, Alberto L. Sangiovanni-Vincentelli, Werner Damm, Thomas A. Henzinger, and Kim G. Larsen. 2018. Contracts for System Design. *Found. Trends Electron. Des. Autom.* 12, 2-3 (2018), 124–400.
[10] Jan A. Bergstra and Cornelis A. Middelburg. 2005. Process algebra for hybrid systems. *Theor. Comput. Sci.* 335, 2-3 (2005), 215–280.
[11] Chiara Bodei, Pierpaolo Degano, Gian-Luigi Ferrari, and Letterio Galletta. 2016. Where Do Your IoT Ingredients Come From?. In *COORDINATION 2016 (LNCS, Vol. 9686)*. Springer, 35–50.
[12] Ningning Chen and Huibiao Zhu. 2023. A process calculus SMrCaIT for IoT. *J. Softw. Evol. Process.* (2023).
[13] Xin Chen, Sriram Sankaranarayanan, and Erika Ábrahám. 2014. Under-approximate flowpipes for non-linear continuous systems. In *FMCAD 2014*. IEEE, 59–66.
[14] Gabriel Ciobanu and Maciej Koutny. 2011. Timed Mobility in process algebra and Petri nets. *J. Log. Algebraic Methods Program.* 80, 7 (2011), 377–391.
[15] Pieter J. L. Cuijpers and Michel A. Reniers. 2005. Hybrid process algebra. *J. Log. Algebraic Methods Program.* 62, 2 (2005), 191–245.
[16] Guillaume Dupont, Yamine Aït Ameur, Neeraj Kumar Singh, and Marc Pantel. 2021. Event-B Hybridation: A Proof and Refinement-based Framework for Modelling Hybrid Systems. *ACM Trans. Embed. Comput. Syst.* 20, 4 (2021), 35:1–35:37.
[17] Goran Frehse, Colas Le Guernic, Alexandre Donzé, Scott Cotton, Rajarshi Ray, Olivier Lebeltel, Rodolfo Ripado, Antoine Girard, Thao Dang, and Oded Maler. 2011. SpaceEx: Scalable Verification of Hybrid Systems. In *CAV 2011 (LNCS, Vol. 6806)*. Springer, 379–395.

[18] Ting Gan, Mingshuai Chen, Yangjia Li, Bican Xia, and Naijun Zhan. 2018. Reachability Analysis for Solvable Dynamical Systems. *IEEE Trans. Autom. Control.* 63, 7 (2018), 2003–2018.
[19] Francesco Gavazzo. 2018. Quantitative Behavioural Reasoning for Higher-order Effectful Programs: Applicative Distances. In *LICS 2018*. ACM, 452–461.
[20] Antoine Girard and George J. Pappas. 2007. Approximation Metrics for Discrete and Continuous Systems. *IEEE Trans. Autom. Control.* 52, 5 (2007), 782–798.
[21] Dimitar P. Guelev, Shuling Wang, and Naijun Zhan. 2017. Compositional Hoare-Style Reasoning About Hybrid CSP in the Duration Calculus. In *SETTA 2017 (LNCS, Vol. 10606)*. Springer, 110–127.
[22] Jifeng He. 1994. From CSP to Hybrid Systems. In *A Classical Mind: Essays in Honour of C. A. R. Hoare*. Prentice Hall International (UK) Ltd., 171–189.
[23] Thomas A. Henzinger. 1996. The Theory of Hybrid Automata. In *LICS 1996*. IEEE Computer Society, 278–292.
[24] C. A. R. Hoare. 1978. Communicating Sequential Processes. *Commun. ACM* 21, 8 (1978), 666–677.
[25] A. Agung Julius, Alessandro D'Innocenzo, Maria Domenica Di Benedetto, and George J. Pappas. 2009. Approximate equivalence and synchronization of metric transition systems. *Syst. Control. Lett.* 58, 2 (2009), 94–101.
[26] Hui Kong, Fei He, Xiaoyu Song, William N. N. Hung, and Ming Gu. 2013. Exponential-Condition-Based Barrier Certificate Generation for Safety Verification of Hybrid Systems. In *CAV (LNCS, Vol. 8044)*. Springer, 242–257.
[27] Gerardo Lafferriere, George J. Pappas, and Sergio Yovine. 2001. Symbolic Reachability Computation for Families of Linear Vector Fields. *J. Symb. Comput.* 32, 3 (2001), 231–253.
[28] Ugo Dal Lago and Maurizio Murgia. 2023. Contextual Behavioural Metrics. In *CONCUR 2023 (LIPIcs, Vol. 279)*. Schloss Dagstuhl - Leibniz-Zentrum für Informatik, 38:1–38:17.
[29] Ivan Lanese, Luca Bedogni, and Marco Di Felice. 2013. Internet of things: a process calculus approach. In *SAC 2013*. ACM, 1339–1346.
[30] Ruggero Lanotte and Massimo Merro. 2018. A semantic theory of the Internet of Things. *Information and Computation* 259 (2018), 72–101.
[31] Jiang Liu, Jidong Lv, Zhao Quan, Naijun Zhan, Hengjun Zhao, Chaochen Zhou, and Liang Zou. 2010. A Calculus for Hybrid CSP. In *APLAS 2010 (LNCS, Vol. 6461)*. Springer, 1–15.
[32] Rupak Majumdar, Nobuko Yoshida, and Damien Zufferey. 2020. Multiparty motion coordination: from choreographies to robotics programs. *Proc. ACM Program. Lang.* 4, OOPSLA (2020), 134:1–134:30.
[33] MathWorks. 2013. *Simulink® User's Guide.* http://www.mathworks.com/help/pdf_doc/simulink/sl_using.pdf.
[34] MathWorks. 2013. *Stateflow® User's Guide.* http://www.mathworks.com/help/pdf_doc/stateflow/sf_ug.pdf.
[35] Robin Milner. 1993. *The Polyadic π-Calculus: A Tutorial.* Springer.
[36] Robin Milner. 1999. *Communicating and Mobile Systems: the π-Calculus.* Cambridge University Press.
[37] Robin Milner, Joachim Parrow, and David Walker. 1992. A Calculus of Mobile Processes, I. *Inf. Comput.* 100, 1 (1992), 1–40.
[38] Robin Milner, Joachim Parrow, and David Walker. 1992. A Calculus of Mobile Processes, II. *Inf. Comput.* 100, 1 (1992), 41–77.
[39] Pablo A Parrilo. 2000. *Structured semidefinite programs and semialgebraic geometry methods in robustness and optimization.* California Institute of Technology.
[40] André Platzer. 2008. Differential Dynamic Logic for Hybrid Systems. *J. Autom. Reason.* 41, 2 (2008), 143–189.
[41] André Platzer. 2012. The Complete Proof Theory of Hybrid Systems. In *LICS*. IEEE Computer Society, 541–550.
[42] André Platzer. 2018. *Logical Foundations of Cyber-Physical Systems.* Springer.
[43] Stephen Prajna and Ali Jadbabaie. 2004. Safety Verification of Hybrid Systems Using Barrier Certificates. In *HSCC (LNCS, Vol. 2993)*. Springer, 477–492.
[44] William C. Rounds. 2004. A Spatial Logic for the Hybrid Pi-Calculus. In *HSCC 2004 (LNCS, Vol. 2993)*. Springer, 508–522.
[45] William C. Rounds and Hosung Song. 2003. The Phi-Calculus: A Language for Distributed Control of Reconfigurable Embedded Systems. In *HSCC 2003 (LNCS, Vol. 2623)*. Springer, 435–449.
[46] Alceste Scalas and Nobuko Yoshida. 2019. Less is more: multiparty session types revisited. *Proc. ACM Program. Lang.* 3, POPL (2019), 30:1–30:29.
[47] D. A. van Beek, Ka L. Man, Michel A. Reniers, Jacobus E. Rooda, and Ramon R. H. Schiffelers. 2006. Syntax and consistent equation semantics of hybrid Chi. *J. Log. Algebraic Methods Program.* 68, 1-2 (2006), 129–210.
[48] D. A. van Beek, Jacobus E. Rooda, Ramon R. H. Schiffelers, Ka L. Man, and Michel A. Reniers. 2005. Relating Hybrid Chi to Other Formalisms. In *IFM Doctoral Symposium 2005 (ENTCS, Vol. 191)*. Elsevier, 85–113.
[49] Peter C. W. van den Brand, Michel A. Reniers, and Pieter J. L. Cuijpers. 2006. Linearization of hybrid processes. *J. Log. Algebraic Methods Program.* 68, 1-2 (2006), 54–104.
[50] Rob J. van Glabbeek and W. P. Weijland. 1996. Branching Time and Abstraction in Bisimulation Semantics. *J. ACM* 43, 3 (1996), 555–600.

[51] J. Wang, V. Magron, and J. Lasserre. 2021. TSSOS: a moment-SOS hierarchy that exploits term sparsity. *SIAM Journal on optimization* 31, 1 (2021), 30–58.

[52] Qiuye Wang, Mingshuai Chen, Bai Xue, Naijun Zhan, and Joost-Pieter Katoen. 2022. Encoding inductive invariants as barrier certificates: Synthesis via difference-of-convex programming. *Information and Computation* 289, Part (2022), 104965.

[53] Shuling Wang, Zekun Ji, Xiong Xu, Bohua Zhan, Qiang Gao, and Naijun Zhan. 2024. Formally Verified C Code Generation from Hybrid Communicating Sequential Processes. In *ICCPS 2024*. IEEE, 123–134.

[54] Hao Wu, Shenghua Feng, Ting Gan, Jie Wang, Bican Xia, and Naijun Zhan. 2024. On Completeness of SDP-Based Barrier Certificate Synthesis over Unbounded Domains. In *FM (2) (LNCS, Vol. 14934)*. Springer, 248–266.

[55] Wanling Xie, Huibiao Zhu, and Xi Wu. 2023. A Timed Calculus with Mobility for Wireless Networks. In *CNIOT 2023*. ACM, 653–657.

[56] Wanling Xie, Huibiao Zhu, and Qiwen Xu. 2021. A process calculus BigrTiMo of mobile systems and its formal semantics. *Formal Aspects Comput.* 33, 2 (2021), 207–249.

[57] Xiong Xu, Shuling Wang, Bohua Zhan, Xiangyu Jin, Jean-Pierre Talpin, and Naijun Zhan. 2022. Unified graphical co-modeling, analysis and verification of cyber-physical systems by combining AADL and Simulink/Stateflow. *Theor. Comput. Sci.* 903 (February 2022), 1–25.

[58] Gaogao Yan, Li Jiao, Shuling Wang, Lingtai Wang, and Naijun Zhan. 2020. Automatically Generating SystemC Code from HCSP Formal Models. *ACM Trans. Softw. Eng. Methodol.* 29, 1 (2020), 4:1–4:39.

[59] Naijun Zhan, Bohua Zhan, Shuling Wang, Dimitar Guelev, and Xiangyu Jin. 2023. A Generalized Hybrid Hoare Logic. arXiv:2303.15020 [cs.LO]

[60] Chaochen Zhou, Ji Wang, and Anders P. Ravn. 1995. A Formal Description of Hybrid Systems. In *Hybrid Systems III: Verification and Control (LNCS, Vol. 1066)*. Springer, 511–530.

## A   PROOFS OF THE STRONG CONGRUENCE

LEMMA A.1. *The following relations are strong bisimulations:*

*(1)* $\{((v\vec{z})(P\|X), (v\vec{z})(Q\|Y)) \mid P \sim Q, X \sim Y, \vec{z} \in \mathcal{N}^*\}$

*(2)* $\{(!P, !Q), ((v\vec{z})(X\|!P), (v\vec{z})(Y\|!Q)) \mid P \sim Q, X \sim Y, \vec{z} \in \mathcal{N}^*\}$

*where $\mathcal{N}$ is the set of names.*

PROOF. (1) Let $\mathcal{S}$ denote the relation. We first consider the case that $(v\vec{z})(P\|X) \xrightarrow{\lambda} A$, where $\lambda$ is a discrete action. We try to find some B such that $(v\vec{z})(Q\|Y) \xrightarrow{\lambda} B$ and $A\mathcal{S}B$. The former transition can be inferred by any of the following rules:

$$\frac{P \xrightarrow{\lambda} A' \quad \lambda \notin \vec{z} \cup \overline{\vec{z}}}{(v\vec{z})(P\|X) \xrightarrow{\lambda} (v\vec{z})(A'\|X)} \qquad \frac{X \xrightarrow{\lambda} A' \quad \lambda \notin \vec{z} \cup \overline{\vec{z}}}{(v\vec{z})(P\|X) \xrightarrow{\lambda} (v\vec{z})(P\|A')}$$

$$\frac{P \xrightarrow{x} F \quad X \xrightarrow{\overline{x}} C}{(v\vec{z})(P\|X) \xrightarrow{\tau} (v\vec{z})(F@C)} \qquad \frac{P \xrightarrow{\overline{x}} C \quad X \xrightarrow{x} F}{(v\vec{z})(P\|X) \xrightarrow{\tau} (v\vec{z})(F@C)}$$

where $\vec{z} \cup \overline{\vec{z}} \triangleq \{z, \overline{z} \mid z \in \vec{z}\}$.

For the first transition, since $P \xrightarrow{\lambda} A'$ and $P \sim Q$, we get $Q \xrightarrow{\lambda} B'$ and $A' \sim B'$. Thus,

$$\frac{Q \xrightarrow{\lambda} B' \quad \lambda \notin \vec{z} \cup \overline{\vec{z}}}{(v\vec{z})(Q\|X) \xrightarrow{\lambda} (v\vec{z})(B'\|X)}$$

For A' and B', we have the following three cases:

**Case 1** A' and B' are processes. Then, $(v\vec{z})(A'\|X)\mathcal{S}(v\vec{z})(B'\|X)$ given that $A' \sim B'$.

**Case 2** Let $A' = (\vec{x}).P'$ and $B' = (\vec{x}).Q'$ be abstractions, where $\mathcal{N}(\vec{x}) \cap (\mathcal{F}(X) \cup \mathcal{N}(\vec{z})) = \emptyset$. Since $A' \sim B'$, $\exists \mathcal{S}'$ such that $A'\mathcal{S}'B'$, i.e., $P'\{\vec{e}/\vec{x}\} = A'\langle \vec{e} \rangle \mathcal{S}' B'\langle \vec{e} \rangle = Q'\{\vec{e}/\vec{x}\}$ for all $\vec{e}$ by definition. Thus, $(v\vec{z})(P'\{\vec{e}/\vec{x}\}\|X)\mathcal{S}(v\vec{z})(Q'\{\vec{e}/\vec{x}\}\|X)$ for all $\vec{e}$. By Definition 5.2 and $\mathcal{N}(\vec{x}) \cap (\mathcal{F}(X) \cup \mathcal{N}(\vec{z})) = \emptyset$, we can get $(v\vec{z})(A'\|X) = (\vec{x}).(v\vec{z})(P'\|X)$ and $(v\vec{z})(B'\|X) = (\vec{x}).(v\vec{z})(Q'\|X)$ and for all $\vec{e}$,

$$\begin{array}{ccc} (\vec{x}).(v\vec{z})(P'\|X)\langle\vec{e}\rangle & & (\vec{x}).(v\vec{z})(Q'\|X)\langle\vec{e}\rangle \\ \| & & \| \\ (v\vec{z})(P'\{\vec{e}/\vec{x}\}\|X) & \mathcal{S} & (v\vec{z})(Q'\{\vec{e}/\vec{x}\}\|X) \end{array}$$

which means

$$\begin{array}{ccc} (v\vec{z})(A'\|X) & & (v\vec{z})(B'\|X) \\ \| & & \| \\ (\vec{x}).(v\vec{z})(P'\|X) & \mathcal{S} & (\vec{x}).(v\vec{z})(Q'\|X) \end{array}$$

by definition.

**Case 3** A' and B' are concretions. Since $A' \sim B'$, $\exists \mathcal{S}'$ such that $A'\mathcal{S}'B'$ and we let $A' = (v\vec{y})\langle \vec{e} \rangle.P'$ and $B' = (v\vec{y})\langle \vec{e} \rangle.Q'$ such that $P'\mathcal{S}'Q'$, where $\mathcal{N}(\vec{y}) \cap (\mathcal{F}(X) \cup \mathcal{N}(\vec{z})) = \emptyset$. By Definition 5.2,

$$\begin{array}{rcl} (v\vec{z})(A'\|X) & = & (v\vec{z}_1, \vec{y})\langle\vec{e}\rangle.(v\vec{z}_2)(P'\|X) \\ (v\vec{z})(B'\|X) & = & (v\vec{z}_1, \vec{y})\langle\vec{e}\rangle.(v\vec{z}_2)(Q'\|X) \end{array}$$

where we assume $\vec{z} = \vec{z}_1, \vec{z}_2$ with $\mathcal{N}(\vec{z}_1) \subseteq \mathcal{N}(\vec{e})$ and $\mathcal{N}(\vec{z}_2) \cap \mathcal{N}(\vec{e}) = \emptyset$. Since $P'\mathcal{S}'Q'$, we get

$$(v\vec{z}_2)(P'\|X) \quad \mathcal{S} \quad (v\vec{z}_2)(Q'\|X)$$

which implies $(v\vec{z})(A'\|X)\mathcal{S}(v\vec{z})(B'\|X)$ by definition.

Let $A = (v\vec{z})(A'\|X)$ and $B = (v\vec{z})(B'\|X)$, then $A\mathcal{S}B$. The proof for the second transition is symmetric.

For the third transition, since $P \xrightarrow{x} F$, $X \xrightarrow{\bar{x}} C$, $P \sim Q$, and $X \sim Y$, we get $Q \xrightarrow{x} F'$, $Y \xrightarrow{\bar{x}} C'$, $F \sim F'$, and $C \sim C'$. Thus, we get the following transition:

$$\frac{Q \xrightarrow{x} F' \quad X \xrightarrow{\bar{x}} C'}{(v\vec{z})(Q\|X) \xrightarrow{\tau} (v\vec{z})(F'@C')}$$

Since $F \sim F'$ and $C \sim C'$, $\exists \mathcal{S}', \mathcal{S}''$ such that $F\mathcal{S}'F'$ and $C\mathcal{S}''C'$. Thus, we can let $F = (\vec{x}).P'$, $F' = (\vec{x}).Q'$, $C = (v\vec{y})\langle\vec{e}\rangle.X'$, and $C' = (v\vec{y})\langle\vec{e}\rangle.Y'$ such that $P'\{\vec{e}'/\vec{x}\} = F\langle\vec{e}'\rangle \mathcal{S}' F'\langle\vec{e}'\rangle = Q'\{\vec{e}'/\vec{x}\}$ for all $\vec{e}'$, which implies $P'\{\vec{e}/\vec{x}\}\mathcal{S}'Q'\{\vec{e}/\vec{x}\}$, and $X'\mathcal{S}''Y'$. Then,

$$(v\vec{z})(F@C) \qquad (v\vec{z})(F'@C')$$
$$\| \qquad \|$$
$$(v\vec{z},\vec{y})(P'\{\vec{e}/\vec{x}\}\|X') \quad \mathcal{S} \quad (v\vec{z},\vec{y})(Q'\{\vec{e}/\vec{x}\}\|Y')$$

Let $A = (v\vec{z})(F@C)$ and $B = (v\vec{z})(F'@C')$, then $A\mathcal{S}B$. The proof for the fourth transition is symmetric.

Now we consider the case that $(v\vec{z})(P\|X) \xrightarrow{\langle\rho_1 \frown \cdots \frown \rho_n, \mathcal{R}\rangle} R$. We try to find some $R'$ such that $(v\vec{z})(Q\|Y) \xrightarrow{\langle\rho_1, \mathcal{R}\rangle} \cdots \xrightarrow{\langle\rho_n, \mathcal{R}\rangle} R'$ and $R\mathcal{S}R'$. The former transition can be inferred by the rule:

$$\frac{P \xrightarrow{\langle\rho', \mathcal{R}'\rangle} P' \quad X \xrightarrow{\langle\rho'', \mathcal{R}''\rangle} X' \quad \overline{\mathcal{R}'} \cap \mathcal{R}'' = \emptyset}{(v\vec{z})(P\|X) \xrightarrow{\langle\rho' \otimes \rho'', \mathcal{R}' \cup \mathcal{R}''\rangle \setminus \vec{z}} (v\vec{z})(P'\|X')}$$

where $\langle\rho' \otimes \rho'', \mathcal{R}' \cup \mathcal{R}''\rangle \setminus \vec{z} = \langle\rho_1 \frown \cdots \frown \rho_n, \mathcal{R}\rangle$. Since $P \xrightarrow{\langle\rho', \mathcal{R}'\rangle} P'$, $X \xrightarrow{\langle\rho'', \mathcal{R}''\rangle} X'$, $P \sim Q$, and $X \sim Y$, we get $Q \xrightarrow{\langle\rho'_1, \mathcal{R}'\rangle} \cdots \xrightarrow{\langle\rho'_n, \mathcal{R}'\rangle} Q'$, $Y \xrightarrow{\langle\rho''_1, \mathcal{R}''\rangle} \cdots \xrightarrow{\langle\rho''_n, \mathcal{R}''\rangle} Y'$, $P' \sim Q'$, and $X' \sim Y'$, where

$$\rho' = \rho'_1 \frown \cdots \frown \rho'_n \text{ and } \rho'' = \rho''_1 \frown \cdots \frown \rho''_n$$

with $|\rho_i| = |\rho'_i| = |\rho''_i|$ for $1 \leq i \leq n$. Then, we can get the transitions

$$(v\vec{z})(Q\|Y) \xrightarrow{\langle\rho'_1 \otimes \rho''_1, \mathcal{R}' \cup \mathcal{R}''\rangle \setminus \vec{z}} \cdots \xrightarrow{\langle\rho'_n \otimes \rho''_n, \mathcal{R}' \cup \mathcal{R}''\rangle \setminus \vec{z}} (v\vec{z})(Q'\|Y')$$

Since $\langle\rho' \otimes \rho'', \mathcal{R}' \cup \mathcal{R}''\rangle \setminus \vec{z} = \langle\rho_1 \frown \cdots \frown \rho_n, \mathcal{R}\rangle$, we get $(\rho'_i \otimes \rho''_i) \setminus \vec{z} = \rho_i$ and $(\mathcal{R}' \cup \mathcal{R}'') \setminus \vec{z} = \mathcal{R}$, i.e.,

$$(v\vec{z})(Q\|Y) \xrightarrow{\langle\rho_1, \mathcal{R}\rangle} \cdots \xrightarrow{\langle\rho_n, \mathcal{R}\rangle} (v\vec{z})(Q'\|Y')$$

Since $P' \sim Q'$ and $X' \sim Y'$, we get $(v\vec{z})(P'\|X')\mathcal{S}(v\vec{z})(Q'\|Y')$. Let $R = (v\vec{z})(P'\|X')$ and $R' = (v\vec{z})(Q'\|Y')$, then $R\mathcal{S}R'$. In summary, $\mathcal{S}$ is a strong simulation and so is its converse: $\mathcal{S}$ is a strong bisimulation.

(2) Let $\mathcal{S}$ denote the relation. We first consider the case that $!P \xrightarrow{\lambda} A$, where $\lambda$ is a discrete action. We try to find some $B$ such that $!Q \xrightarrow{\lambda} B$ and $A\mathcal{S}B$. Based on rules [Rep], [Rep'], [Par], and [Sync], transition $!P \xrightarrow{\alpha} A$ can be inferred by any of the following rules:

$$\frac{P \xrightarrow{\lambda} A'}{!P \xrightarrow{\lambda} A'\|!P} \quad \frac{P \xrightarrow{x} F \quad P \xrightarrow{\bar{x}} C}{!P \xrightarrow{\tau} F@(C\|!P)} \quad \frac{P \xrightarrow{\bar{x}} C \quad P \xrightarrow{x} F}{!P \xrightarrow{\tau} (F\|!P)@C}$$

For the first transition, since $P \xrightarrow{\lambda} A'$ and $P \sim Q$, we get $Q \xrightarrow{\lambda} B'$ and $A' \sim B'$. Thus,

$$\frac{Q \xrightarrow{\lambda} B'}{!Q \xrightarrow{\lambda} B'\|!Q}$$

Similar to the proof for the first transition in (1), we consider $A'$ and $B'$ could be processes (**Case 1**), abstractions (**Case 2**), and concretions (**Case 3**), and for each case, we can prove $(A'\|\,!P)\mathcal{S}(B'\|\,!Q)$. Let $A = A'\|\,!P$ and $B = B'\|\,!Q$, then $A\mathcal{S}B$.

For the second transition, $P \xrightarrow{x} F$, $P \xrightarrow{\bar{x}} C$, and $P \sim Q$, we get $Q \xrightarrow{x} F'$, $Q \xrightarrow{\bar{x}} C'$, $F \sim F'$, and $C \sim C'$. Thus, we can get

$$\frac{Q \xrightarrow{x} F' \quad Q \xrightarrow{\bar{x}} C'}{!Q \xrightarrow{\tau} F'@(C'\|\,!Q)}$$

Since $F \sim F'$ and $C \sim C'$, $\exists \mathcal{S}', \mathcal{S}''$ such that $F\mathcal{S}'F'$ and $C\mathcal{S}''C'$. Thus, we can let $F = (\vec{x}).P'$, $F' = (\vec{x}).Q'$, $C = (\nu\vec{y})\langle\vec{e}\rangle.P''$, and $C' = (\nu\vec{y})\langle\vec{e}\rangle.Q''$ such that $P'\{\vec{e'}/\vec{x}\} = F\langle\vec{e'}\rangle \mathcal{S}' F'\langle\vec{e'}\rangle = Q'\{\vec{e'}/\vec{x}\}$ for all $\vec{e'}$, which implies $P'\{\vec{e}/\vec{x}\}\mathcal{S}'Q'\{\vec{e}/\vec{x}\}$, and $P''\mathcal{S}''Q''$. By Definition 5.3 and 5.2, we get

$$F@(C\|\,!P) = (\vec{x}).P'@(\nu\vec{y})\langle\vec{e}\rangle.(P''\|\,!P) = (\nu\vec{y})(P'\{\vec{e}/\vec{x}\}\|P''\|\,!P)$$

and $F'@(C'\|\,!Q) = (\vec{x}).Q'@(\nu\vec{y})\langle\vec{e}\rangle.(Q''\|\,!Q) = (\nu\vec{y})(Q'\{\vec{e}/\vec{x}\}\|Q''\|\,!Q)$. Since $P'\{\vec{e}/\vec{x}\} \sim Q'\{\vec{e}/\vec{x}\}$, $P'' \sim Q''$, and $P \sim Q$, we get

$$\begin{array}{ccc} F@(C\|\,!P) & & F'@(C'\|\,!Q) \\ \| & & \| \\ (\nu\vec{y})(P'\{\vec{e}/\vec{x}\}\|P''\|\,!P) & \mathcal{S} & (\nu\vec{y})(Q'\{\vec{e}/\vec{x}\}\|Q''\|\,!Q) \end{array}$$

Let $A = F@(C\|\,!P)$ and $B = F'@(C'\|\,!Q)$, then $A\mathcal{S}B$. The proof for the third transition is similar.

Now we consider the case that $!P \xrightarrow{\langle\rho_1\frown\cdots\frown\rho_n,\mathcal{R}\rangle} !P'$. We try to find some $!Q'$ such that

$$!Q \xrightarrow{\langle\rho_1,\mathcal{R}\rangle} \cdots \xrightarrow{\langle\rho_n,\mathcal{R}\rangle} !Q'$$

The former transition can be inferred by the following rule:

$$\frac{P \xrightarrow{\langle\rho,\mathcal{R}\rangle} P' \quad \overline{\mathcal{R}} \cap \mathcal{R} = \emptyset}{!P \xrightarrow{\langle\rho,\mathcal{R}\rangle} !P'}$$

where $\rho = \rho_1 \frown \cdots \frown \rho_n$. Since $P \xrightarrow{\langle\rho,\mathcal{R}\rangle} P'$ and $P \sim Q$, $Q = Q_1 \xrightarrow{\langle\rho_1,\mathcal{R}\rangle} Q_2 \xrightarrow{\langle\rho_2,\mathcal{R}\rangle} \cdots \xrightarrow{\langle\rho_{n-1},\mathcal{R}\rangle} Q_n = Q'$ and $P' \sim Q'$. According to rule [REP']:

$$\frac{Q_i \xrightarrow{\langle\rho_i,\mathcal{R}\rangle} Q_{i+1} \quad \overline{\mathcal{R}} \cap \mathcal{R} = \emptyset}{!Q_i \xrightarrow{\langle\rho_i,\mathcal{R}\rangle} !Q_{i+1}}$$

we get the transitions $!Q \xrightarrow{\langle\rho_1,\mathcal{R}\rangle} \cdots \xrightarrow{\langle\rho_n,\mathcal{R}\rangle} !Q'$. Since $P' \sim Q'$, we get $!P'\mathcal{S}!Q'$ by definition.

For pairs $((\nu\vec{z})(X\|\,!P), (\nu\vec{z})(Y\|\,!Q))$, we can get similar result in the same way. In summary, $\mathcal{S}$ is a strong simulation and so is its converse: $\mathcal{S}$ is a strong bisimulation. □

THEOREM 6.3 (CONGRUENCE).
- *If $A \sim B$ then $\lambda A + M \sim \lambda B + M$.*
- *If $P \sim Q$ then (1) $[B].P + M \sim [B].Q + M$, (2) $(\nu x)P \sim (\nu x)Q$, (3) $P\|R \sim Q\|R$, (4) $!P \sim !Q$, (5) $(\nu\vec{x})\langle\vec{e}\rangle.P \sim (\nu\vec{x})\langle\vec{e}\rangle.Q$, and (6) $\{\vec{e}_0 \mid \dot{\vec{v}} = \vec{e}\&B, \mathcal{R}\}(\vec{y}).P + M \sim \{\vec{e}_0 \mid \dot{\vec{v}} = \vec{e}\&B, \mathcal{R}\}(\vec{y}).Q + M$.*
- *If $P\{\vec{e}/\vec{x}\} \sim Q\{\vec{e}/\vec{x}\}$ for all $\vec{e}$, then $(\vec{x}).P \sim (\vec{x}).Q$.*

Proof. It can be demonstrated that

$$
\begin{aligned}
\mathcal{S}_1 &= \{(\lambda\mathsf{A} + \mathsf{M}, \lambda\mathsf{B} + \mathsf{M})\} \cup \{(\mathsf{P}, \mathsf{Q}) \mid \mathsf{P} \sim \mathsf{Q}\} \\
\mathcal{S}_2 &= \{(([B].\mathsf{P} + \mathsf{M}, [B].\mathsf{Q} + \mathsf{M}), (\mathsf{P}, \mathsf{Q}) \mid \mathsf{P} \sim \mathsf{Q}\} \\
\mathcal{S}_3 &= \{(\{\vec{e}_0 \mid \dot{\vec{v}} = \vec{e}\&B, \mathcal{R}\}(\vec{y}).\mathsf{P} + \mathsf{M}, \{\vec{e}_0 \mid \dot{\vec{v}} = \vec{e}\&B, \mathcal{R}\}(\vec{y}).\mathsf{Q} + \mathsf{M})\} \\
&\quad \cup \{(\widetilde{\pi}.\mathsf{P}, \widetilde{\pi}.\mathsf{Q}), (\mathsf{P}, \mathsf{Q}) \mid \mathsf{P} \sim \mathsf{Q}, \widetilde{\pi} \in C\} \\
\mathcal{S}_4 &= \{((vx)\mathsf{P}, (vx)\mathsf{Q}) \mid \mathsf{P} \sim \mathsf{Q}\}
\end{aligned}
$$

where $C$ is the set of continuous prefixes, are strong bisimulations, which are enough to prove

$$
\begin{aligned}
\lambda\mathsf{A} + \mathsf{M} &\sim \lambda\mathsf{B} + \mathsf{M} \\
[B].\mathsf{P} + \mathsf{M} &\sim [B].\mathsf{Q} + \mathsf{M} \\
\{\vec{e}_0 \mid \dot{\vec{v}} = \vec{e}\&B, \mathcal{R}\}(\vec{y}).\mathsf{P} + \mathsf{M} &\sim \{\vec{e}_0 \mid \dot{\vec{v}} = \vec{e}\&B, \mathcal{R}\}(\vec{y}).\mathsf{Q} + \mathsf{M} \\
(vx)\mathsf{P} &\sim (vx)\mathsf{Q}
\end{aligned}
$$

for $\mathsf{A} \sim \mathsf{B}$ and $\mathsf{P} \sim \mathsf{Q}$. According to Lemma A.1, we can get

$$
\begin{aligned}
\mathsf{P} \| \mathsf{R} &\sim \mathsf{Q} \| \mathsf{R} \\
!\mathsf{P} &\sim !\mathsf{Q}
\end{aligned}
$$

Assume $\mathsf{P} \sim \mathsf{Q}$, then $\exists \mathcal{S}$ such that $\mathsf{P}\mathcal{S}\mathsf{Q}$ and thus $(v\vec{x})\langle\vec{e}\rangle.\mathsf{P}\mathcal{S}(v\vec{x})\langle\vec{e}\rangle.\mathsf{Q}$ by definition, which implies

$$(v\vec{x})\langle\vec{e}\rangle.\mathsf{P} \sim (v\vec{x})\langle\vec{e}\rangle.\mathsf{Q}$$

Assume $\mathsf{P}\{\vec{e}/\vec{x}\} \sim \mathsf{Q}\{\vec{e}/\vec{x}\}$ for all $\vec{e}$, then $\exists \mathcal{S}_{\vec{e}}$ . $\mathsf{P}\{\vec{e}/\vec{x}\}\mathcal{S}_{\vec{e}}\mathsf{Q}\{\vec{e}/\vec{x}\}$. It can be demonstrated that $\mathcal{S} = \bigcup_{\vec{e}} \mathcal{S}_{\vec{e}}$ is a strong bisimulation and $(\vec{x}).\mathsf{P}\mathcal{S}(\vec{x}).\mathsf{Q}$ by definition, thus $(\vec{x}).\mathsf{P} \sim (\vec{x}).\mathsf{Q}$. In summary, $\sim$ is a congruence in the hybrid $\pi$-calculus. □

## B BARRIER CERTIFICATES

This appendix provides the formal definition of a barrier certificate and explains how to use it to compute a differential invariant. For a deeper dive into this topic, we suggest referring to [26, 43, 52, 54]. The following definition and theorem are taken from [26]. We slightly modify their formulations for our setting.

*Definition B.1 (Hybrid Automaton).* A hybrid automaton is a tuple $\mathcal{H} = \langle L, X, E, R, G, I, F \rangle$, where
- $L$ is a finite set of locations;
- $X \subseteq \mathbb{R}^n$ is the continuous state space. The hybrid state space of the system is denoted by $\mathcal{X} = L \times X$ and a state is denoted by $(l, x) \in \mathcal{X}$;
- $E \subseteq L \times A \times L$ is a set of discrete transitions, where $A$ is a set of actions;
- $G : E \mapsto 2^X$ is a guard mapping over discrete transitions;
- $R : E \times X \mapsto 2^X$ is a reset mapping over discrete transitions;
- $I : L \mapsto 2^X$ is an invariant mapping.

THEOREM B.2 (HYBRID-EXP CONDITION BARRIER CERTIFICATE). *[26, Theorem 2] Given a hybrid automaton $\mathcal{H} = \langle L, X, E, R, G, I, F \rangle$, the initial set $\mathcal{X}_0$ and the unsafe set $\mathcal{X}_u$ of $\mathcal{H}$, then, for any given set of constant real numbers $S_\lambda = \{\lambda_l \in \mathbb{R} \mid l \in L\}$ and any given set of constant non-negative real numbers $S_\gamma = \{\gamma_e \in \mathbb{R}_+ \mid e \in E\}$, if there exists a set of continuously differentiable functions $\{\varphi_l(x) \mid l \in L\}$ such that, for all $l \in L$ and $e \in E$, the following formulas hold:*

$$\forall x \in Init(l).\ \varphi_l(x) \leq 0. \tag{BC-1}$$

$$\forall x \in I(l).\ \langle\nabla\varphi_l, f_l\rangle - \lambda_l \varphi_l(x) \leq 0. \tag{BC-2}$$

$$\forall x \in G(e), \forall x' \in R(e, x).\ \gamma_{l,l'}\varphi_l(x) - \varphi_{l'}(x') \geq 0. \tag{BC-3}$$

$$\forall x \in Unsafe(l).\ \varphi_l(x) > 0. \tag{BC-4}$$

where $\nabla$ is the gradient operator, $\langle \cdot, \cdot \rangle$ is the dot product, $Init(l)$ and $Unsafe(l)$ denote respectively the initial set and the unsafe set at location $l$, then the safety property is satisfied by $\mathcal{H}$.

The functions $\{\varphi_l(x) \mid l \in L\}$ are referred to as barrier certificates associated to the hybrid automaton $\mathcal{H}$. If such barrier certificates exist, then the set $\Omega$ defined by

$$\Omega \triangleq \bigcap_{l \in L} \{(l, x) \mid \varphi_l(x) \leq 0\}$$

is a differential invariant that witnesses the safety of $\mathcal{H}$. This is ensured by the conditions outlined in the above theorem. Intuitively, condition BC-1 guarantees that the initial states are included in $\Omega$, while BC-4 ensures that the unsafe states are excluded. Additionally, BC-2 ensures that trajectories can not enter the unsafe region during continuous evolution within each location, while BC-3 guarantees the same property for discrete transitions between successive locations. By assuming that the barrier certificates are of certain polynomial (linear in our case) forms, the constraints can be translated into sum-of-squares optimization problems and be solved as semidefinite programmings [39].

To obtain the differential invariant in the proof of Proposition 7.1, we apply the above theorem to the hybrid automaton $\mathcal{H}$ in Fig. (4). Since $\mathcal{H}$ only has one location, we only need to compute one function $\varphi$ (the subscript $l$ is omitted). We employ the tools TSSOS [51] and Mosek [3] to formulate and solve the constraints. After trying different values, we set $\gamma_e = 1$ for each $e \in E$ and $\lambda = 0.25$. The following barrier certificate is obtained, keeping five decimal places during computation:

$$\begin{aligned}\varphi \triangleq{}& 0.12386 \cdot p_1 + 0.60533 \cdot v_1 - 0.00588 \cdot a_1 - 8.19308 \cdot c_1 \\ &+ 0.12017 \cdot p_2 + 0.58482 \cdot v_2 - 0.03074 \cdot a_2 + 0.64709 \cdot c_2 - 0.40900.\end{aligned}$$

The numerical results can be verified by Mathematica. The code can be found via the link: https://anonymous.4open.science/r/TrainBisimulation-EE75.